\documentclass[journal,12pt, onecolumn,draftclsnofoot]{IEEEtran}

\usepackage[english]{babel}
\usepackage{amsthm}
\theoremstyle{definition}
\newtheorem{theorem}{Theorem}
\newtheorem{lemma}[theorem]{Lemma}
\usepackage[usenames, dvipsnames]{color}
\usepackage{tabularx,ragged2e,booktabs,caption}
\usepackage{cite}
\usepackage{amsmath}
\DeclareMathOperator{\diag}{diag}  
\usepackage{amssymb}
\usepackage{latexsym}
\usepackage{float}
\usepackage{epsfig}
\usepackage{subfigure,epstopdf,graphicx,array,algorithm}
\usepackage[]{graphicx}
\usepackage{algpseudocode}
\usepackage{cancel}
\usepackage{moreverb} 
\usepackage[colorlinks,bookmarksopen,bookmarksnumbered,citecolor=red,urlcolor=red]{hyperref}
\usepackage{graphicx}
\usepackage{cite,amsmath,amsthm,amssymb,float,latexsym,epsfig,subfigure,epstopdf,array}
\usepackage{physics} 
\usepackage{mathtools} 
\usepackage[flushleft]{threeparttable}
\usepackage{makecell,booktabs}
\usepackage{marvosym} 
\usepackage{footnote}
\usepackage{epstopdf}
\usepackage{mathtools,amssymb}

\makeatletter
\renewcommand{\fnum@figure}{Fig. \thefigure}
\makeatother
\usepackage{tikz}

\usepackage{array}
\usepackage{multirow}

\begin{document}
	
	\title{{On Performance of Integrated Satellite HAPS Ground Communication: Aerial IRS Node vs Terrestrial IRS Node  }}
	\author{
		\IEEEauthorblockN{Parvez Shaik, Kamal Kishore Garg, Praveen Kumar Singya, Vimal Bhatia, \IEEEmembership{Senior~Member,~IEEE}, and Mohamed-Slim Alouini, \IEEEmembership{Fellow, IEEE}}
		\thanks{Parvez Shaik is with the Department of Electrical and Computer Engineering, Texas A \& M University at QATAR (e-mail: parvez.shaik@qatar.tamu.edu). This work was done by him while he was with IIT Indore.}
		\thanks{K. K. Garg is with the School of Technology,
			Pandit Deendayal Energy University, Gandhinagar, Gujarat
			(e-mail:kamal.garg@sot.pdpu.ac.in) } 	\thanks{V. Bhatia is with the Department of Electrical Engineering, Indian Institute of Technology Indore, Indore-453552, India (e-mail:  vbhatia@iiti.ac.in)}
		\thanks{P. K. Singya is with the Department of Electrical and Electronics Engineering, ABV-IIITM Gwalior, India (email: praveens@iiitm.ac.in)}  \thanks{M.-S. Alouini is with the Computer, Electrical, and Mathematical Science and Engineering (CEMSE) Division, King Abdullah University	of Science and Technology (KAUST), Thuwal 23955-6900, Saudi Arabia (e-mail: slim.alouini@kaust.edu.sa)}
	}
	\maketitle
	\begin{abstract}
		With a motive of ubiquitous connectivity over the globe with enhanced spectral efficiency, intelligent reflecting surfaces (IRS) integrated satellite-terrestrial communications is a topic of research interest in an infrastructure-deficient remote terrains. In line with this vision, this paper entails the performance analysis of satellite-terrestrial networks leveraging both aerial and terrestrial IRS nodes, with the support of high altitude platforms over diverse fading channels including shadowed Rician, Rician, and Nakagami-$m$ fading channels. The merits of IRS in enhancing spectral efficiency is analyzed  through closed-form expressions of outage probability and ergodic rate. Further, the average symbol error rate analysis for the higher-order quadrature amplitude modulation (QAM) schemes such as hexagonal QAM, rectangular QAM, cross QAM, and square QAM  is performed. Practical constraints like antenna gains, path loss, and link fading are considered  to characterize the satellite terrestrial links. Finally, a comparison between the high-altitude platforms based IRS node and terrestrial IRS nodes is performed and various insights are drawn under various fading scenarios and path loss conditions. This paper contribute towards understanding and potential implementation of IRS-integrated satellite-terrestrial networks for efficient and reliable communication.
	\end{abstract}
	
	\section{Introduction}
	
	The  current research community is progressively moving towards sixth generation (6G) communications to achieve unprecedented seamless  connectivity with ultra-low latency. However, the current terrestrial network poses fundamental challenges in achieving the 6G key performance indicators (KPIs) \cite{chaoub20216g}. The current terrestrial infrastructure is vulnerable to natural disasters and network densification inevitably causing serious economic and environmental concerns as well as energy crunch \cite{giordani2020non}. To circumvent these limitations, in the blueprint of beyond 5G and 6G future telecommunications, the non-terrestrial networks (NTNs)  play an important role in achieving the ambitious KPIs which were not fulfilled by the terrestrial infrastructures alone \cite{huang2019performance}.  With the recent technological advancement and  evolution, third generation partnership project (3GPP) has  enforced this potential vision and detailed it in technical reports \cite{3gpp2019study}. 
	
	In general, the NTNs are satellites,  high-altitude platforms (HAPs), and  unmanned aerial vehicles (UAVs) which are distinguished  based on the height, frequency, coverage area, and propagation of flight operation \cite{ye2022non}.   HAPs act as data center between the UAVs and the satellites. Aerial networks are capable of ensuring unmatched connectivity and services even in the remote inaccessible terrains. However, in practice, aerial networks performance is hampered due to shadowing effects, path losses,  antenna misalignment, power constraints, and hardware limitations \cite{liu2018space}. With the technological advances, an innovative solution in the form of intelligent-reflecting surfaces (IRS) has emerged which plays a crucial role  in realizing the system level KPIs for terrestrial and NTNs  for 6G communications.
	IRS is also known as reconfigurable intelligent surfaces,  develops an intelligent smart radio propagation environment to control the planar wavefront of the incident signal  to increase the signal strength \cite{holloway2012overview}. IRS is a group of planar software defined advanced microelectrical mechanical systems (MEMS) and metamaterials which are engineered electronically to reconfigure the wave steering to realize a desired transformation  in time-varying wireless propagation environments \cite{alghamdi2020intelligent}. Hence,  IRS is very appealing to integrate with satellite-terrestrial communications to meet the 6G KPIs.
	
	Research community in the academia and industry presented their studies on prospects of potential improvements in the wireless systems with the integration of IRS. Authors in \cite{tekbiyik2020reconfigurable} investigated the performance of IRS based satellite communications and analyzed the error rates. Xu \emph{et al.}  \cite{xu2021intelligent}, addressed the secured cooperative communications with IRS for  satellite-terrestrial links. Reference \cite{li2021reconfigurable} performed the secrecy maximization for high mobility drone communication with IRS. To address the prospects of coverage area improvements, the authors of \cite{ ibrahim2021exact, tian2022enabling} proposed  satellite-IRS-user links deployments. The potential use cases and  challenges associated with IRS integration with aerial platforms are presented in \cite{alfattani2021aerial, bariah2021ris, ramezani2022toward}. In \cite{tekbiyik2022reconfigurable},  authors studied IRS assisted non-terrestrial and inter-planetary communications, along with the performance of IRS assisted NTNs. A comparative study is presented with IRS and cooperative relays systems  in  \cite{bjornson2019intelligent, boulogeorgos2020performance, di2020reconfigurable} to explore the potentiality of IRS over relays.  Impact of co-channel interference is investigated in \cite{ sikri2021reconfigurable} for a terrestrial IRS  based system. Dolas \emph{et.al}  \cite{dolas2022performance} studied  satellite-terrestrial system in terms of outage probability and error rate analysis by modeling satellite to relay link  with Shadowed-Rician distributed and cascaded relay-IRS-destination link is Rayleigh distributed.  In \cite{dong2022intelligent}, authors  investigated an IRS-aided integrated terrestrial-satellite network system by deploying IRS to assist both the terrestrial and satellite systems. In majority of the literature discussed, authors considered the IRS in the terrestrial link of NTNs considered. However, analysis of IRS in NTNs as a aerial node versus a terrestrial node is missing. To bridge this gap, we attempt to address the issue by considering IRS as a aerial node and a terrestrial node for a integrated satellite HAP ground communication.

	On the other hand, the signal characteristics especially modulation schemes play a vital role for a reliable power-efficient high-speed communications. In this prospective, the bandwidth and power efficient higher-order quadrature amplitude modulation (QAM) schemes have gained significant attraction. Depending on the constellations, family of QAMs includes square QAM (SQAM), rectangular QAM (RQAM), cross QAM (XQAM), and hexagonal QAM (HQAM) \cite{singya2021survey}. In the past, majority of the works on the average symbol error rate (ASER) of higher-order QAM schemes are performed  in RF communications, optical wireless communications (OWC) (especially ultra-violet communications and free space optics), and in mixed RF/OWC systems \cite{Parvez_D2D_Access_2019, parvez2019impact, shaik2019performance, garg2019performance, 10.1117/1.OE.59.1.016106, singya2021performance, garg2022performance}. Thus it is of high interest to perform ASER analysis of higher order QAM schemes for the system employing IRS.

	\subsection{Motivation}
	
	One of the primary driving motive behind the development of 6G is the pursuit of end-to-end (e2e) connectivity that offers extreme agility, flexibility, low cost, low latency, and power-efficiency. While terrestrial infrastructures are inadequate to meet these ambitious requirements, a promising solution lies in the integration of hybrid terrestrial and aerial networks. This hybrid structure can ensure uninterrupted service even during natural disasters and facilitate highly secure military applications with cost-effectiveness. However, non-terrestrial communications face challenges such as atmospheric turbulence, weather conditions, and power limitations. To overcome these obstacles and achieve the desired Key Performance Indicators (KPIs) in 6G, IRS emerges as a potential technology. IRS plays a crucial role in enhancing the performance of aerial nodes and terrestrial nodes for NTNs.
	
	In this context, the present work focuses on two distinct system models:
	
	\begin{itemize}
		\item System Model 1 (Line-of-Sight - LoS): This model involves three nodes: a satellite ($\text{S}$) as the information source, an IRS mounted on a High-Altitude Platform (HAP) with N-elements ($\text{H}_\text{I}$), and an end user ($\text{U}$) as the destination node. The channel link $\text{S} \rightarrow \text{H}_\text{I}$ exhibits dominant LoS propagation, modeled by Rician fading. The link $\text{H}_\text{I} \rightarrow \text{U}$ is modeled using SR fading to capture the impact of non-terrestrial-terrestrial  links.
		\item 	System Model 2 (Non-Line-of-Sight - NLoS): This model involves four nodes: a satellite ($\text{S}$) as the information source, a HAP acting as a decode-and-forward (DF) relay ($\text{H}_\text{R}$), a terrestrial IRS with N-elements ($\text{I}_\text{N}$), and an end user ($\text{U}$) as the destination node. The $\text{S} \rightarrow \text{H}_\text{R}$ link assumes an LoS path modeled by Rician fading ($h{s}$). The link $\text{H}\text{I} \rightarrow \text{I}_\text{N}$ is modeled with SR fading ($h_i$), while the terrestrial link $\text{I}_\text{N} \rightarrow \text{U}$ is modeled using generalized Nakagami-$m$ flat fading channels.  This configuration is designed to handle NLoS scenarios effectively by employing a terrestrial relay node with IRS deployment to forward data from the HAP to terrestrial users.
	\end{itemize}
	For both system models, it is assumed that all nodes have a single antenna. Additionally, a communication-oriented software ensures precise control of phase-shifts for incident signals on the IRS, enabling coherent/constructive signal combining at the end user ($\text{U}$) to maximize the e2e signal-to-noise ratio (SNR). This coherent combining technique enhances the overall system performance.
	\subsection{Contribution}
	Considering the above system models, the main contributions of this work are as follows:
	\begin{itemize}
		\item The cumulative distribution function (CDF) of the e2e SNR, and the exact closed-form expression of outage probability for  both system models is derived. The effect of antenna beam, path loss, and satellite beam gains on both the systems performance is illustrated. 
		\item  ASER performance analysis of various  QAM schemes such as HQAM, RQAM, and XQAM  for both system models is analyzed and useful inferences are drawn.
		\item The closed-form expressions of the ergodic rate for both system models are derived to determine the maximum possible transmission rate by the systems. Further, the improvement in the capacity with an increase in IRS elements under  detrimental channel fading and propagation losses is validated.
		\item Finally, the impact of path losses ( antenna beam, rain, attenuation, free-space pathloss), satellite beam angle, and IRS elements shown on both systems performance along with various fading channel conditions. 
	\end{itemize}

	\noindent {\textbf{Notations:} 
		Column vectors are denoted by bold lowercase letters; 
		%
		%
		Squared Frobenius norm is denoted by $||\cdot||^2$. 
		Nakagami-$m$ distribution with fading severity $m$ and variance ${\sigma^2_m}$ is denoted by $\text{Nak}(m,{\sigma^2_m})$.
		Complex Gaussian distribution with mean 0, variance $\sigma^2$ is denoted by $\mathcal{CN}(0,\sigma^2)$.
		Confluent Hypergeometric function (HF) of first kind and Gauss HF are represented by ${}_{1}{{F}_{1}}(a,b,c)$ and ${}_{2}{{F}_{1}}(a,b,c,d)$, respectively. 
		Probability density function (PDF) and cumulative distribution function (CDF) are given by $f(\cdot)$ and $F(\cdot)$, respectively.
		Generalized Marcum Q-function is given by $Q_m(\cdot,\cdot)$. 
		%
		%
		Statistical expectation operator and variance are denoted by $\mathrm{E}\{(\cdot)\}$
		and $\mathrm{Var}\{(\cdot)\}$, respectively.
		Modified Bessel function of first kind with order $\vartheta$ is denoted as $I_{\vartheta}(\cdot)$.
		Bessel function of first kind with order $\varrho$ is denoted as $J_{\varrho}(\cdot)$.
		%
		Exponential integral function is denoted as $\bf{Ei}(\cdot)$.
		Gamma function is represented by $\Gamma(\cdot)$
		%
		Upper incomplete gamma function with parameters \{a,b\} is represented as $\Gamma(a,b)$. 
		Finally, $ G_{p,q}^{m,n} \left( \begin{matrix}
			a_1, a_2,...,a_p \\
			b_1, b_2,...,b_q
		\end{matrix} \bigg| \begin{matrix}
			z
		\end{matrix} \right)$ is the Meijer-G function with $0\leq m\leq q$ and $0\leq n\leq p$, where $m$, $n$, $p$, and $q$ are integers.
	} 

\section{Statistical characterization of Channel}
\subsection{Channel Modeling}
This section focuses on the statistical	properties and the channel models of both the satellite and terrestrial links, present in the considered system models. For the design and performance analysis of the real-time operation based satellite-terrestrial mobile communications, shadowed Rician (SR) model is employed to model both  narrowband and wideband communciations \cite{abdi2003new}. 
\begin{figure*}[htp]
	\centering
	\subfigure[ System Model 1.\label{SY1}]{\includegraphics[width=1.8in]{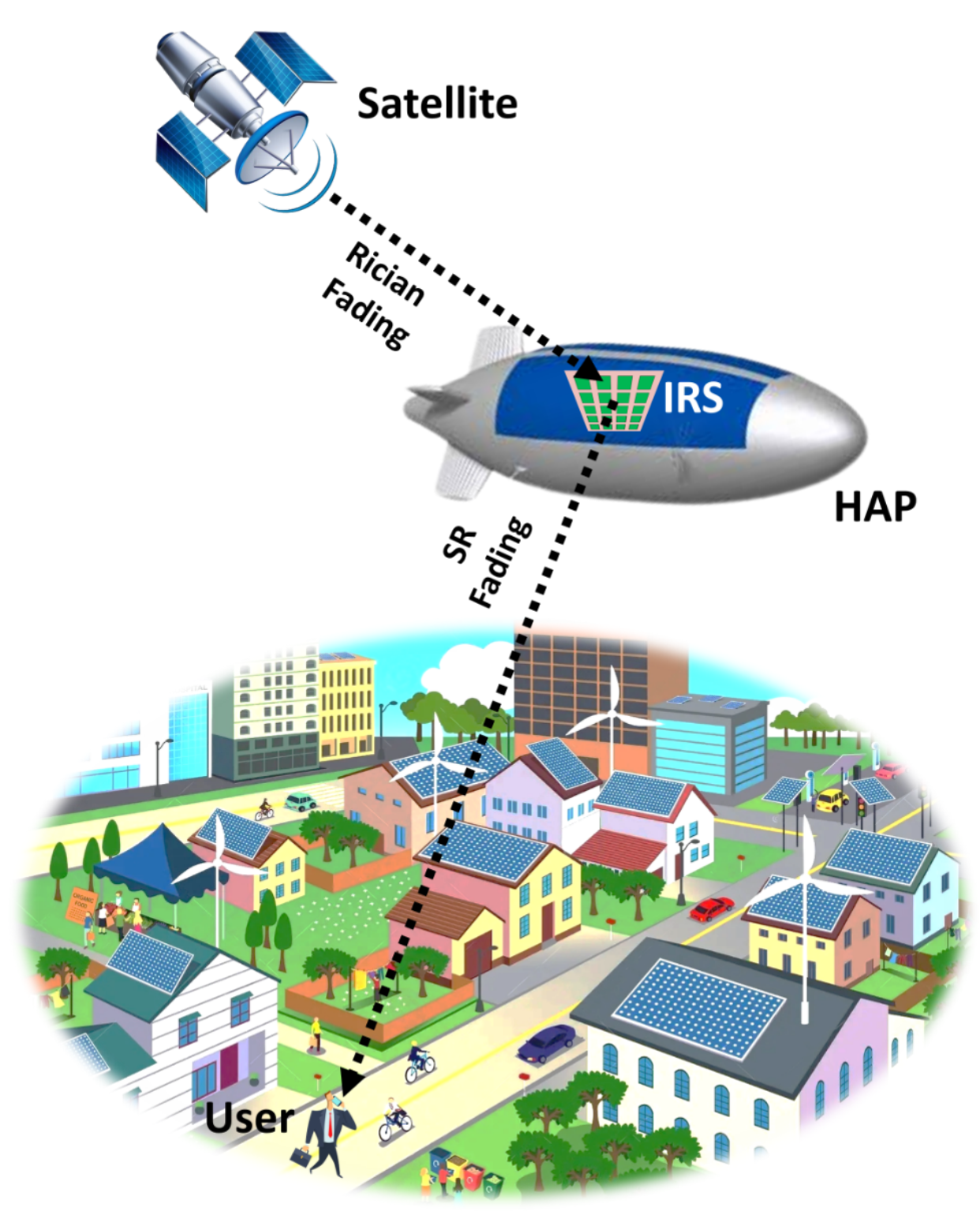}}\quad
	\subfigure[System Model 2. \label{SY2}]{\includegraphics[width=1.8in]{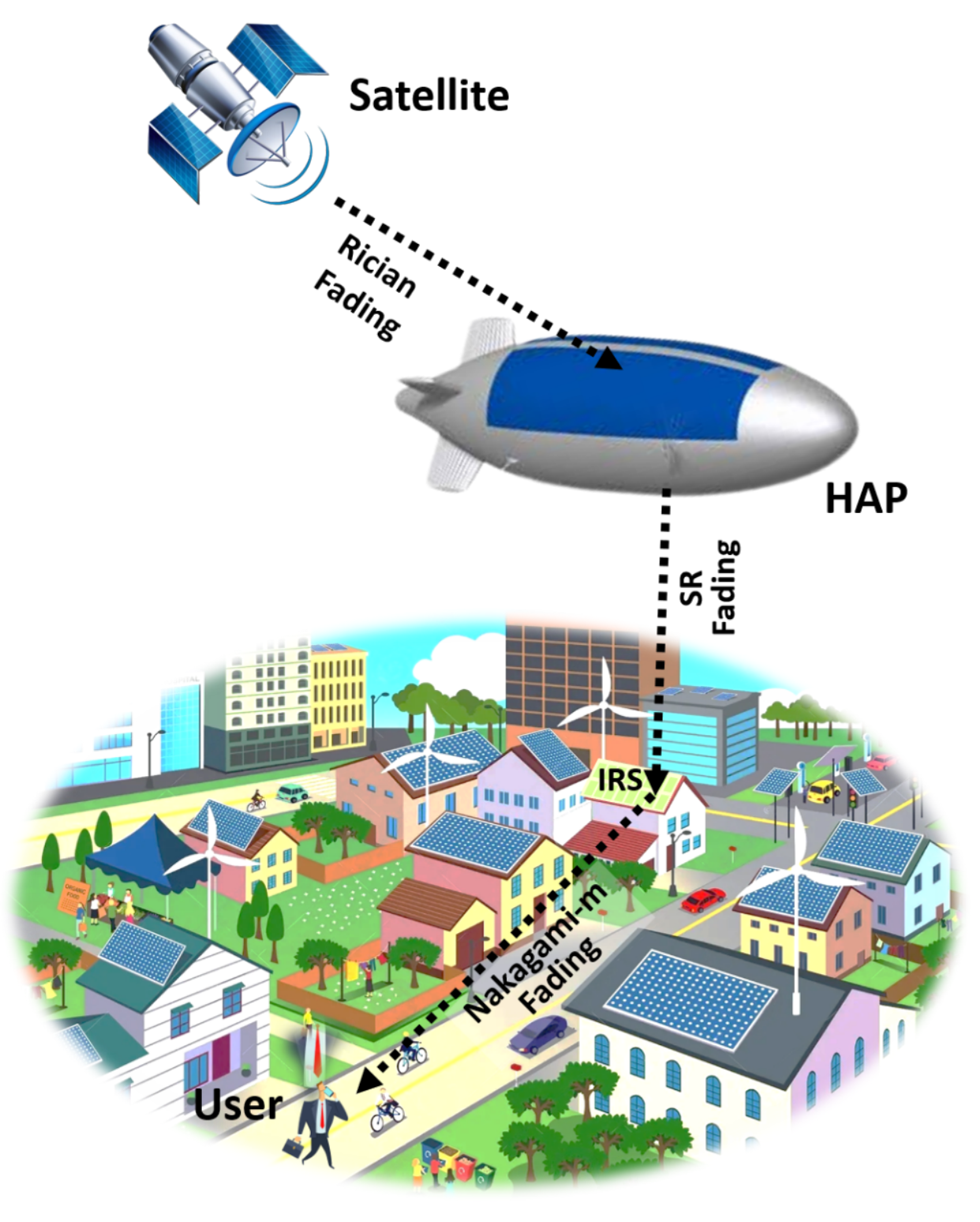}}\quad
	\caption{Integrated satellite HAPs ground communication system models}
	\label{SM}
\end{figure*}

The statistical characterization of the channels is as follows:

\subsubsection{Rician Fading Channel}	
The channel gain $\gamma$ is  noncentral-
$\chi^2$ distributed and the probability density function (PDF) of the  Rician fading channel is given as \cite[eq. 5]{suraweera2009performance},
\begin{align} \label{PDFR}
	f_{\gamma_{i}}(\gamma)={{(K+1)e^{-K}}\over {\bar{\gamma}_{i}}}e^{-{(K+1)\gamma}\over {\bar{\gamma}_{i}}}I_{0}\left(2\sqrt{{K(K+1)\gamma}\over{\bar{\gamma}_{i}}}\right), \quad {\gamma} \geq 0
\end{align}
where $K$ is the Rician K-factor,  $\bar{\gamma}_{i}=\frac{\Omega_i}{\sigma^2_N}$,  $\Omega_i$ is the average fading power of $i^{th}$ link, and $\sigma^2_N $ is the the power of the additive white Gaussian noise (AWGN) component. The CDF is given as \cite[(2.3-65)]{proakis1995digital},
\begin{align} \label{CDFR}
	F_{\gamma_{i}}(\gamma)=1-\mathcal{Q}_1\left(\sqrt{(2K)},  \sqrt{\frac{2(K+1)\gamma}{\bar{\gamma}_{i}} } \right), \quad {\gamma} \geq 0
\end{align}
where $S_R=\mu_R$ and local-mean scattered power is given as $\sigma_R^2=\sqrt{\frac{S_R^2}{2 K}}$.	The $k^{th}$ moment of the Rician fading channels is given as  $	\mathrm{E}(x^k)= \left(2\sigma^2_R\right)^{\frac{k}{2}} \Gamma\left(1+\frac{k}{2}\right){}_1{F_1}\left({-\frac{k}{2}},1,-K \right)$.

\subsubsection{Shadowed Rician Fading Channel}
The components of SR fading channels are considered to be independent and identically distributed as $	g_i=Z\exp(j\zeta)+A\exp(j\psi)$ \cite{abdi2003new},
where independent stationary random processes $Z$ and
$A$ are the amplitudes of the LoS and the scatter components, which follow Nakagami-$m$ and Rayleigh distributions, respectively. In addition, $\zeta$ is the deterministic phase of the
LoS component, while $\psi$ is the stationary random phase
process with uniform distribution over $[0, 2\pi)$, which are
independent of Z and A. The shadowed Rician fading channel PDF of $|g_i|^2$ is given as  \cite[eq. (6)]{abdi2003new}, 
\begin{align}\label{SR}
	f_{|g_i|^2}=\alpha {e^{ - \beta x}}{}_1{F_1}\left({m;1;\delta x} \right),\quad x \geq 0
\end{align}
where $\alpha=\frac{\left[\frac{2bm}{2bm+\Omega}\right]^m}{2b}$, $\beta=\frac{1}{2b}$, and $\delta=\frac{\Omega}{2b[2bm+\Omega]}$, $\Omega$ is the average power of the LoS component($Z$) given by $\Omega=\mathrm{E}[Z^2]$. Also, the average power of the scatter component is given by $2b=\mathrm{E}[A^2]$, $m=\frac{(\mathrm{E}[Z^2])^2}{\mathrm{Var}[Z^2]}$, Nakagami-$m$ parameter corresponding to severity of fading, and Rician parameter $K=\frac{\Omega}{2b} $.  For different types of shadowing, the values of $(b,m,\Omega)$ are given in \cite[Table III]{abdi2003new}. The $k^{th}$ moment of the shadowed Rician fading channels is given as \cite[eq. (5)]{abdi2003new}
\begin{align}
	\mathrm{E}(x^k) =& \left( {{2b_0 m}\over {2b_0 m + \Omega}} \right)^m (2b_0)^{k/2} \Gamma \left( {{k}\over {2}} + 1 \right) {}_2F_1 \left( {{k}\over {2}} + 1,\, m,\, 1,\, {{\Omega }\over {2b_0 m + \Omega }} \right) \cr \noalign{\vskip 4pt} &\quad k = 0,\, 1,\, 2,\, \ldots 
\end{align}
The  closed-form PDF expression of $||\textbf{g}_i||^2=\sum_{i=1}^{N}|g_i|^2$ is given as \cite[eq. 26]{huang2019performance}
\begin{align}
	f_{||g_i||^2}=\frac{\alpha}{\Gamma(N)} x^{N-1}{e^{ - \beta x}}{}_1{F_1}\left({Nm;N;\delta x} \right), \quad x \geq 0
\end{align}
\subsubsection{Nakagami-$m$ fading channel}
The PDF of the Nakagami-$m$ flat fading channel is given as \cite[eq.(2.3-67)]{proakis1995digital}
\begin{align}\label{NPDF}
	f_{g_i}(x)= \frac{2x^{2m_{g}-1}}{\Gamma(m_{g})\zeta_{g}^{m_{g}}}\exp\left(\frac{-x^{2}}{\zeta_{g}}\right),\quad x\geq 0
\end{align}
where $\zeta_g=\frac{\sigma^2_g}{m_g}$. The $k^{th}$ moment of the Nakagami-$m$ fading channels is given as $	\mathrm{E}(x^k)=\frac{\Gamma{(m+\frac{k}{2})}}{\Gamma{(m)}}\left( \frac{{\sigma^2}}{m} \right)^\frac{k}{2}$  \cite[eq.(2.3-70)]{proakis1995digital},
\subsubsection{Pathloss Modeling}
In this subsection,  path losses are discussed  which are essential in  modeling the  realistic scenarios of satellite-terrestrial communication 
\subsubsection*{ $\text{S} \rightarrow \text{H}$ Link} The considered path loss  includes the effects of antenna beam and free path loss and is given as \cite{shuai2022transmit}
\begin{align}
	PL_{\text{SH}} = { \lambda_\text{c}\sqrt{G_\text{S}G_\text{H}} \over 4\pi d_{\text{SH}} \sqrt{K_\text{B}T_{\text{N}_0} B_{\text{N}_0}} },
\end{align}
where $\lambda_\text{c}$ is the operating carrier wavelength,  $G_\text{H}$ is the gain of HAP, $ d_{\text{SH}}$ is the distance between the GEO stationary satellite and the HAP, $K_\text{B}$ is the Boltzmann constant ($ 1.38\times10^{-23}   $ joule/K), $T_{\text{N}_0} $ is the system noise temperature ($300$ K), and $B_{\text{N}_0}$ is the receiver noise bandwidth ($20$ MHz). Further,   $G_\text{S}$ is the satellite beam gain which is given as
%
%
%
%
\begin{align}
	G_\text{S} = G_\text{Max} \left({ J_1(u_\text{H}) \over 2 u_\text{H}} + 36 {J_3(u_\text{H}) \over  u_\text{H}^3} \right)^2,
\end{align}
where
$G_\text{Max}$ is the maximum satellite beam gain.  Also,  $u_\text{H}=2.07123{\sin(\phi_\text{SH})\over\sin(\phi_\text{3dB})}$ with $\phi_\text{SH}$ being the angle between \text{H} and $S$ and $\phi_\text{3dB} $ is the 3dB satellite beam angle.
\subsubsection*{Other links}
In the system model \figurename{ \ref{SY1}}, for  the $\text{H}_\text{I} \rightarrow \text{U}$ link and in system model \figurename{\ref{SY2}}, for  $\text{H}_\text{R} \rightarrow \text{I}_\text{N}$ link, the path loss are modeled as $PL_\text{(uv)}[\text{dB}] = G_\text{Tx}+G_\text{Rx}-L_\text{FSPL}-L_\text{Rain}-L_\text{Atm.}-L_\text{oth}$,
where $\{uv\} \in \{\text{H}_\text{I}  \text{U}, \text{H}_\text{R}  \text{I}_\text{N}\}$,  $G_\text{Tx}$ is the transmitter antenna gain (in dB), $G_\text{Rx}$ is the receiver antenna gain (in dB), $L_\text{Rain}$ is the rain induced loss (in dB/km), $L_\text{Atm.}$ is the gaseous atmospheric absorption loss, $L_\text{oth}$ are other miscellaneous losses (in dB). Further, $L_\text{FSPL}$ is the free-space
path loss (in dB),  given as $L_\text{FSPL} =92.45+20\log_{10}(f_c)$+20$\log_{10}(d_\text{uv})$ with $f_c$ being the carrier frequency, and $d_\text{uv}$ is the distance between the nodes $\text{u}$ and $\text{v}$.

%
\subsubsection*{ $\text{I}_\text{N} \rightarrow \text{U}$ link}
In system model  \figurename{ \ref{SY2}},  the path loss corresponding to the $\text{I}_\text{N} \rightarrow \text{U}$ link is modeled as: $PL_{(\text{I}_\text{N} \rightarrow \text{U})}[\text{dB}] =  
40 \log_{10}(d_{\text{I}_\text{N}\text{U}})- 10 \log_{10}(Gt_\text{IRS})- 10 \log_{10}(Gr_\text{U}) - 20 \log_{10}(h_t) $ $
- 20 \log_{10}(h_r)  $,
where $d_{\text{I}_\text{N}\text{U}}$ is the distance between $\text{I}_\text{N}$ and $\text{U}$,  $Gt_\text{IRS}$ is the gain induced by IRS network, $Gr_\text{U}$ is the gain of $\text{U}$, $h_t$ is the height of the building upon which IRS is mounted,  and $h_r$ is the height at which the UE is situated.

\section{Performance of System Model 1}
For the analysis, the channels are  block-faded, where within the coherence interval, the channels are considered to be static and frequency flat. $\text{S} \rightarrow \text{H}_\text{I}$ link is represented as $h_i$  whereas the $\text{H}_\text{I} \rightarrow \text{U}$ link is represented as $g_i$. Further,  $0< i\leq N$ indicates indices of the IRS element. As considered, when a signal is incident on the IRS, the phase-shifts of the incident signals are controlled perfectly,  whereas the signals reflected from the IRS surface are attenuated as per the reflective coefficient. Hence, $\text{S} \rightarrow \text{H}_\text{I} \rightarrow \text{U}$ channel links  for the $i^{th}$ IRS element is given in the polar form as
\begin{align}
	h_i=\alpha_i e^{j\theta},  \quad
	g_i=\beta_i e^{j\phi},
\end{align}
where $\alpha_i=|h_i|$ and $\beta_i=|g_i|$ are the  channel amplitudes,  respectively, and $\theta_i$ and $\phi_i$ are the phases. The PDF of  $|\alpha_i|^2$  and $|\beta_i|^2$ are given in    \cite{abdi2003new}.	
In the analysis, perfect knowledge of the channel phases of $h_i$ and $g_i$ for $i=1,2,...,N$ at the IRS is assumed, which corresponds to the best scenario in terms of system operation and yields a performance benchmark for practical applications. The signal received at the destination is given as
\begin{align}
	y= \sqrt{P_t}\textbf{g}^T {\xi}  \textbf{h} x+n,
\end{align}
where $P_t$ is the transmit power, $\textbf{h}$ and $\textbf{g}$ are the channel vectors given as $\textbf{h}=[h_1, h_2, ..., h_N]^T$ $\textbf{g}=[g_1, g_2, ..., g_N]^T$, respectively. Further, $ {\xi}=\diag[|\kappa_1|e^{-j\Phi_1}, |\kappa_2|e^{-j\Phi_2},...,|\kappa_N|e^{-j\Phi_N}]$ is a matrix of IRS meta-surface induced complex valued reflection coefficient with attenuation coefficient $\kappa\in[0,1]$ and phase shift $\Phi\in[0,2\pi]$.  Also, $x$ is the symbol transmitted with $\mathrm{E}[|x|^2]=1$, and $n$ is the additive white Gaussian noise (AWGN) at $\text{U}$ with $n \sim C\mathcal{N}(0, \sigma^2_N)$. The received signal at $\text{U}$ can be written as
\begin{align}
	y&= \sqrt{P_t}\left[\sum_{i=1}^{N} h_i \kappa_ie^{-j\Phi_i}  g_i \right] x+n = \sqrt{P_t}\left[\sum_{i=1}^{N} \alpha_i \beta_i \kappa_i e^{-j(\Phi_i-\theta_i-\phi_i)}   \right] x+n, 
\end{align}
Hence, the SNR at $\text{U}$ is obtained as
\begin{align}
	\gamma&= \bar{\gamma}\abs{\sum_{i=1}^{N} \alpha_i \beta_i \kappa_i e^{-j(\Phi_i-\theta_i-\phi_i)} }^2, 
\end{align}
where the average SNR is $\bar{\gamma}=\frac{P_t}{\sigma^2_N}$. To maximize the SNR at $\text{U}$, reflection coefficient induced by $\text{H}_\text{I}$ is chosen optimally such that constructive interference arises and hence, $\kappa_i=1 \forall i$ and $\Phi_i=\theta_i+\phi_i \forall i$ \cite{basar2019wireless}. The optimal maximized SNR at $U$ is given as
\begin{align} \label{a1}
	\gamma&= \bar{\gamma}\abs{\sum_{i=1}^{N} \alpha_i \beta_i }^2.
\end{align}

\subsection*{Performance Metrics:}

\subsubsection{Outage Probability}
A system is in outage if the instantaneous SNR ($\gamma$) of the e2e link falls
bellow a threshold SNR ($\gamma_{th}$).  For conventionally large number of IRS elements, the outage probability can be expressed as
\begin{align}
	P_{out}=\Pr(\gamma<\gamma_{th}). 
\end{align}

\begin{lemma}\label{l1}
	For sufficiently large number of IRS elements, the CDF of e2e link is given as
	\begin{align} \label{CDF1}
		F_{\gamma_{e2e_{1}}}(\gamma)=1-Q_{\frac {1}{2}}\left ({\frac { {\mu_\gamma }}{\sigma_\gamma },\frac {\sqrt {\gamma}}{\sqrt {\bar {\gamma }}\sigma_\gamma }}\right), \quad  \gamma > 0
	\end{align}
\end{lemma}	
\begin{proof}
	Given in Appendix \ref{OP1}.  \qedhere
\end{proof}
\subsubsection{ASER Analysis}
In this subsection,  ASER analysis of the system model 1 for various  modulation schemes is performed. 	The generalized ASER expression for a digital modulation scheme by using the CDF approach \cite[eq. (17)]{parvez2019aser} is given as 
\begin{align}\label{SEP}
	P_{s}(e)=-\int_{0}^{\infty}P'_{s}(e|\gamma)F_{\Lambda_{e2e}}(\gamma)d\gamma, 
\end{align}
%
%
%


\subsubsection*{Hexagonal QAM}
For M-ary HQAM scheme the conditional SEP expression over the AWGN channel is defined as \cite{Garg_2021, singya2021performance}
\begin{align}\label{HSEP}
	P_{s}(e|\gamma)&=H_aQ(\sqrt{\alpha_{h} \gamma})+\frac{2}{3}H_{c}Q^2\bigg(\sqrt{\frac{2\alpha_{h} \gamma}{3}}\bigg)-2H_bQ(\sqrt{\alpha_{h} \gamma})Q\bigg(\sqrt{\frac{\alpha_{h}\gamma}{3}}\bigg),
\end{align}
where the parameters $H_a$, $H_b$, and $\alpha_{h}$ for irregular HQAM are defined in \cite{singya2021survey}.
\begin{lemma}\label{l3}
	The generalized ASER expression for HQAM is given as
		\begin{align}\label{HGE}
			{{P}_{s_1}}^{HQAM}&\approx{1\over 2}\big( {{H}_{a}}-H_b \big) + \frac{{{H}_{b}}}{3} -\frac{{{H}_{b}}}{2} -  \frac{2{{H}_{b}}\alpha_{h}}{9\pi } \frac{3}{2\alpha_{h}} {}_{2}{{F}_{1}}\Big( 1,1,\frac{3}{2},\frac{1}{2} \Big)
			+\frac{{{H}_{b}}\alpha_{h} }{2\sqrt{3}\pi }  \frac{3}{2\alpha_{h}} 
			\nonumber\\&\times
			\left( {}_{2}{{F}_{1}}\Big( 1,1,{3 \over 2},\frac{1}{2} \Big)+{}_{2}{{F}_{1}}\Big( 1,1,\frac{3}{2},\frac{1}{4} \Big) \right)
			+A_2 \bigg( \alpha^\nu  \bigg(  {\Gamma(\mu+\nu)\over \mu}\bigg( \frac{1}{2}\sqrt{\frac{\alpha_{h} }{2\pi }}\big( {{H}_{b}}-H_a \big) 
			\nonumber\\&\times
			\Bbb{F}_2(\mu,\alpha,\beta_{1}) - \frac{{{H}_{b}}}{3}\sqrt{\frac{\alpha_{h} }{3\pi }} 
			\Bbb{F}_2(\mu,\alpha,\beta_{2})+
			\frac{{{H}_{b}}}{2}\sqrt{\frac{\alpha_{h} }{6\pi }} \Bbb{F}_2(\mu,\alpha,\beta_{3}) \bigg) + \sum\limits_{n} 
			\frac{ \Gamma(\mu_2+\nu)}{\mu_2 }
			\nonumber\\&\times
			\Bbb{F}_2(\mu_2,\alpha,\beta_{4}) \bigg( \frac{2{{H}_{b}}\alpha_{h}}{9\pi } {\left(\frac{\alpha_{h}}{3}\right)}^{n} -  \frac{{{H}_{b}}\alpha_{h} }{2\sqrt{3}\pi } 
			\left\{ { \left(\frac{\alpha_{h}}{2} \right)}^{n} + { \left(\frac{\alpha_{h}}{6} \right)}^{n}  \right\} \bigg) \bigg)  \bigg),
		\end{align}		
\end{lemma}
wherein  $A_2=\sum_{k_1=0}^{\infty} e^{-\alpha_{1}^{2}/2}{1 \over k_1!}\left(\alpha_{1}^{2}\over2\right)^{k_1} {1 \over \Gamma(\frac{1}{2}+k_{1})}$, $\nu=\frac{1}{2}+k_1$, $\alpha=\frac { 1} {2{\bar {\gamma_u }}\sigma_{\gamma_u}^2 }$, $\mu=\frac{1}{2}$, $\beta_1={{\frac{\alpha_{h}   }{2}}}$, $\beta_2={{\frac{\alpha_{h}   }{3}}}$, $\beta_3={{\frac{\alpha_{h}   }{6}}}$,  $\mu_2=n+1$,  and  $\beta_4={{\frac{2 \alpha_{h} }{3}}}$.
\begin{proof}
	Given in Appendix \ref{H}.  \qedhere
\end{proof}
\subsubsection*{Rectangular QAM}
Over the AWGN channel, the conditional SEP of RQAM scheme is given as \cite[eq. (14)]{parvez2019aser}
\begin{align}\label{RCSEP}
	{\mathrm{P}_{s}}^{RQAM}(e|\gamma )& =2\Big[ {R_1}Q\big( a_r \sqrt{\gamma } \big)+{R_2}Q\big( b_r \sqrt{\gamma } \big) -2{R_1}{R_2}Q\big( a_r \sqrt{\gamma } \big)Q\big( b_r \sqrt{\gamma } \big) \Big],
\end{align}
where ${R_1}=1-\frac{1}{{{M}_{I}}}$,  ${R_2}=1-\frac{1}{{{M}_{Q}}}$, $a_r =\sqrt{\frac{6}{( {{M}_{I}}^{2}-1 )+( M_{Q}^{2}-1){{d_r }^{2}}}} $ and $b_r =d_r a_r $,  wherein $M_{I}$ and $M_{Q}$ are the number of in-phase and quadrature-phase constellation points, respectively. Also, $d_r =\frac{d_{Q}}{d_{I}}$, where $d_{I}$ and $d_{Q}$ indicate the in-phase and quadrature decision distances, respectively. 
\begin{lemma}\label{l4} 
	The generalized ASER expression for RQAM is given as 
			\begin{align} \label{RGE}
				{{P}_{s_1}}^{RQAM}&\approx I_R +\frac{a_r b_r {R_1}{R_2}}{\pi}  \beta_3^{-1}  \bigg\{ {}_{1}{{F}_{1}}\left( 1,\frac{3}{2},{B_r \over   \beta_3  }   \right ) + {}_{1}{{F}_{1}}\left(1, 1,\frac{3}{2},{A_r \over  \beta_3  }  \right ) \bigg\}
				\nonumber\\&
				%
				%
				+ A_2 \bigg( \alpha^\nu  \bigg(  {\Gamma(\mu+\nu)\over \mu}\bigg(  \frac{a_r{R_1}( {R_2}-1 )}{\sqrt{2\pi  }} \Bbb{F}_2(\mu,\alpha,\beta_{1})	+\frac{b_r {R_2}( {R_1}-1)}{\sqrt{2\pi}}
				\Bbb{F}_2(\mu,\alpha,\beta_{2})
				\nonumber\\&
				-
				\sum\limits_{n}  \frac{a_r b_r {R_1}{R_2}}{\pi}  	\frac{ \Gamma(\mu_2+\nu)}{\mu_2 }\Bbb{F}_2(\mu_2,\alpha,\beta_{3})
				\left( B_r^{n} + A_r^{n}  \right)  \bigg)  \bigg),
			\end{align}
\end{lemma}
where $I_R=-a_r{R_1}( {R_2}-1 ) - b_r {R_2}( {R_1}-1) $, $A_r=\frac{{{a_r }^{2}}}{2}$, $B_r =\frac{{{b_r }^{2}}}{2}$ $\nu=\frac{1}{2}+k_1$; $\alpha=\frac { 1} {2{\bar {\gamma_u }}\sigma_{\gamma_u}^2 }$, $\mu=\frac{1}{2}$, $\beta_1=A_r$, $\beta_2=B_r$, $\mu_2=n+1$, and where $\beta_3=\frac{( {{a_r  }^{2}}+{{b_r }^{2}} ) }{2} $.	SQAM is a special case of RQAM which can be obtained by taking $M_I=M_Q=\sqrt{M}$ and $d_{IQ} = 1$.
\begin{proof}
	Given in Appendix \ref{R}.  \qedhere
\end{proof}
\subsubsection*{Cross QAM}
The conditional SEP of XQAM scheme over AWGN channel is given as \cite[eq. (53)]{SADHWANI201763}
%
		\begin{align}
			&{\mathrm{P}_{s}}^{X}\big( e|\gamma  \big)=A_{n}Q\bigg( \sqrt{\frac{2\gamma }{\alpha_{x}}} \bigg)+\frac{8}{M_{x}N_{x}}\bigg\{\sum\limits_{l}
			Q\bigg(2l\sqrt{\frac{2\gamma}{\alpha_{x}}}\bigg)+Q\bigg(\frac{M_{x}-N_{x}}{2}\sqrt{\frac{2\gamma}{\alpha_{x}}}\bigg)-2\sum\limits_{l}
			\nonumber\\&\times
			Q\bigg(2l\sqrt{\frac{2\gamma}{\alpha_{x}}}\bigg)Q\bigg( \sqrt{\frac{2\gamma }{\alpha_{x}}} \bigg)
			-Q\bigg( \sqrt{\frac{2\gamma }{\alpha_{x}}} \bigg)Q\bigg(\frac{M_{x}-N_{x}}{2}\sqrt{\frac{2\gamma}{\alpha_{x}}}\bigg)\bigg\}
			+k_{x}Q^2{\bigg(\sqrt{\frac{2\gamma}{\alpha_{x}}}\bigg)},
		\end{align}
where $A_{n}=4-2\frac{M_{x}+N_{x}}{M_{x}N_{x}}$,$\sum\limits_{l}=\sum_{l=1}^{\frac{M_{x}-N_{x}}{4}-1}$ $k_{x}=4-4\frac{M_{x}+N_{x}}{M_{x}N_{x}}+\frac{8}{M_{x}N_{x}}$, $\alpha_{x}=\frac{2}{3}(\frac{31M_{x}N_{x}}{32}-1)$, $M_{x}$, and $N_{x}$ are the number of columns and rows corresponding to RQAM. 
\begin{lemma}\label{l5}
	The generalized ASER expression for XQAM is given as
			\begin{align}\label{XGE}
				P_{s_1}^X&\approx\frac{-1}{2}\mathbb{A}_X +\frac{2}{{M_{x}N_{x}}} 
				+ \frac{16}{M_{x}N_{x}} \sum\limits_{l} \frac{l}{\pi \alpha_{x} } A_{x_3}^{-1} \biggl( {}_{1}{{F}_{1}}\Big( 1,\frac{3}{2},\frac{1}{A_{x_3} \alpha_{x}} \Big)
				+{}_{1}{{F}_{1}}\Big( 1,\frac{3}{2},\frac{4 l^2}{A_{x_3} \alpha_{x}} \Big) \biggr)
				\nonumber\\&
				-2\frac{A_{x_2}}{\pi\alpha_{x}} \beta_4^{-1} \bigg( {}_{1}{{F}_{1}}\Big( 1,\frac{3}{2},\frac{A_{x_1}^2}{\beta_4 \alpha_{x}} \Big)
				+{}_{1}{{F}_{1}}\Big( 1,\frac{3}{2},\frac{1}{\beta_4 \alpha_{x}} \Big) \bigg)
				+\frac{k_x}{\pi\alpha_{x}} {\beta_5}^{-1} 
				{}_{1}{{F}_{1}}\Big( 1,\frac{3}{2},\frac{1}{\beta_5 \alpha_{x}} \Big)   \bigg)
				\nonumber\\&
				+A_2 \bigg( \alpha^\nu  \bigg(  {\Gamma(\mu+\nu)\over \mu}\bigg( \frac{\mathbb{A}_X}{2\sqrt{\pi \alpha_{x}}} \Bbb{F}_2(\mu,\alpha,\beta_{1}) 
				-\frac{A_{x_2}}{\sqrt{\pi\alpha_{x}}} \Bbb{F}_2(\mu,\alpha,\beta_{2})  \bigg)  
				+ \sum\limits_{n}  \frac{ \Gamma(\mu_2+\nu)}{\mu_2}  \bigg(   \frac{-16}{M_{x}N_{x}}
				\nonumber\\&\times
				\sum\limits_{l}  \frac{l}{\pi \alpha_{x}} 
				\Bbb{F}_2(\mu_2,\alpha,\beta_{3})  \left( { \left( \frac{1}{\alpha_{x}} \right) }^{n}  + {\left( \frac{4l^2}{\alpha_{x}}  \right)}^{n}  \right) -2\frac{A_{x_2}}{\pi\alpha_{x}} \left( { \left( \frac{1}{\alpha_{x}} \right) }^{n}  + {\left( \frac{A_{x_1}^2}{\alpha_{x}}  \right)}^{n}  \right)
				\nonumber\\&\times
				\Bbb{F}_2(\mu_2,\alpha,\beta_{4})  	
				-\frac{k_x}{\pi\alpha_{x}} \Bbb{F}_2(\mu_2,\alpha,\beta_{5}) { \left( \frac{1}{\alpha_{x}} \right) }^{n}  \bigg) \bigg)   \bigg),
			\end{align}
\end{lemma}
where $A_{x_{1}}={\frac{M_{x}-N_{x}}{2}}$, $A_{x_{2}}=\frac{M_{x}-N_{x}}{M_{x}N_{x}}$, and $A_{x_{3}}=\frac{4l^2+1}{\alpha_{x}}$. Further $\nu=\frac{1}{2}+k_1$, $\alpha=\frac { 1} {2{\bar {\gamma_u }}\sigma_{\gamma_u}^2 }$, $\mu_1=\frac{1}{2}$, $\beta_1=\frac{1}{\alpha_{x}} $,  $\beta_2=A_{x_1}^2\frac{1}{\alpha_{x}} $, $\mu_2=n+1$, $\beta_3= A_{x_3}$, $\beta_4=\frac{1+A_{x_1}^2}{\alpha_{x}} $, and $\beta_5= \frac{2}{\alpha_{x}}$.
\begin{proof}
	Given in Appendix \ref{X}.  \qedhere
\end{proof}
\subsubsection{Ergodic Rate Analysis}
Ergodic rate is defined as the expected value of the instantaneous mutual information between the source and the receiver. It indicates the maximum rate attained by the system. The ergodic rate of the system model \figurename { \ref{SY1}} can be expressed as
\begin{align}
	C_R &=\mathrm{E}\left( \frac{1}{2} \log_2(1+ \gamma_{e2e})\right)
	=\mathrm{E}\left( \frac{1}{2} \log_2(1+\gamma)\right)
	= {1\over{2\ln2}} \int_{0}^{\infty} {{1-F_{\gamma_{e2e}}(\gamma)}\over 1+\gamma} d\gamma.\label{Er}
\end{align}

\begin{lemma}\label{l6}
	Ergodic rate of the considered system is given as
	\begin{align}
		C_{R_ 1} &= \sum_{k_1=0}^{\infty} e^{-\alpha_{1}^{2}/2}{1 \over k_1!}\left(\alpha_{1}^{2}\over2\right)^{k_1} {1 \over \Gamma(\frac{1}{2}+k_{1})}{1\over{2\ln2}} { G_{3,1}^{2,3} \left( \begin{matrix}
				{  { 1} \over {2{\bar {\gamma }}\sigma_{\Lambda}^2}} 
			\end{matrix} \bigg| \begin{matrix}
				0,1 \\ 0,0, \frac{1}{2}+k_{1}
			\end{matrix} \right) }.  \label{er1}
	\end{align}
\end{lemma}
\begin{proof}
	Given in Appendix \ref{ER1}.
\end{proof}

\section{Performance of System Model 2} 
In this system model, it is assumed that the direct link between the $\text{S}$ and $\text{U}$ is not available due to severe blockage effects. $\text{S}$ communicate to $\text{U}$ through a DF $\text{H}_\text{R}$ and $\text{I}_\text{N}$.
In the first time slot, $\text{S}$ transmit the information to $\text{H}_\text{R}$  through an LoS path. The signal received at the $ \text{H}_\text{R}$ from $\text{S}$  is given as
\begin{align}
	y_{r}=\sqrt{P_s}h_s x+n_{s},
\end{align}
where $P_s$ is the transmit power at $\text{S}$,  $x$ is the transmitted  symbol with $\mathrm{E}[|x|^2]=1$  and $n_s$ is the AWGN of the  $\text{S} \rightarrow \text{H}_\text{R}$ link with $ n \sim C\mathcal{N}(0, \sigma^2)$. The instantaneous SNR of the $\text{S} \rightarrow \text{H}_\text{R}$ link is expressed as
\begin{align}
	y_u= \sqrt{P_h}\textbf{g}_{\bf{s}}^T \mathrm{\zeta}  \textbf{h}_{\bf{s}} x+n,
\end{align}
where $P_h$ is the transmit power at  $\text{H}_\text{R}$, $\textbf{h}_{\bf{s}}$ and $\textbf{g}_{\bf{s}}$ are the channel vectors corresponding to shadowed Rician fading and Nakagami-$m$ fading and are denoted as $\textbf{h}_{\bf{s}}=[h_{s_1}, h_{s_2}, ..., h_{s_N}]^T$ and $\textbf{g}_{\bf{s}}=[g_{s_1}, g_{s_2}, ..., g_{s_N}]^T$, respectively. $ \mathbb{\zeta}=\diag[|\varrho_1|e^{-j\Phi_1}, |\varrho_2|e^{-j\Phi_2},...,|\varrho_N|e^{-j\Phi_N}]$ is a matrix of IRS meta-surface induced complex valued reflection coefficient with attenuation coefficient $\varrho\in[0,1]$ and phase shift $\Phi\in[0,2\pi]$.
In our analysis, we assume perfect knowledge of the channel phases of $h_{s_{i}}$ and $g_{s_{i}}$ for $i=1,2,...,N$ at the $\text{I}_\text{N}$, which corresponds to the best scenarios in terms of system operation and yields a performance benchmark for practical applications. The received signal can be re-written as
\begin{align}
	y_u&= \sqrt{P_h}\left[\sum_{i=1}^{N} h_{s_i} \varrho_{s_i} e^{-j\Phi_{s_i}}  g_{s_i} \right] x+n = \sqrt{P_h}\left[\sum_{i=1}^{N} \lambda_i \kappa_{s_i} \varrho_{s_i} e^{-j(\Phi_{s_i}-\theta_{s_i}-\phi_{s_i})}   \right] x+n, 
\end{align}
The SNR at $\text{U}$ is given as
\begin{align}
	\gamma_u&= \bar{\gamma_u}\abs{\sum_{i=1}^{N} \lambda_{s_i} \kappa_{s_i} \varrho_{s_i} e^{-j(\Phi_i-\theta_{s_i}-\phi_{s_i})} }^2,
\end{align}
where $\bar{\gamma_u}=\frac{P_h}{\sigma^2}$. To maximize the SNR at $\text{U}$, reflection coefficient induced by $\text{I}_\text{N}$ must be chosen optimally such that constructive interference increases and hence, $\varrho_{s_i}=1 $ $ \forall i$ and $\Phi_{s_i}=\theta_{s_i}+\phi_{s_i}$ $ \forall i$. The optimal maximized SNR at $\text{U}$ is given as
\begin{align} \label{a2}
	\gamma_u&= \bar{\gamma_u}\abs{\sum_{i=1}^{N} \lambda_{s_i} \kappa_{s_i} }^2.
\end{align}
The e2e SNR of the $\text{S} \rightarrow \text{H}_\text{R} \rightarrow \text{I} \rightarrow \text{U} $ link is given as
\begin{align} \label{a3}
	\gamma_{su}&=  \min\left( \gamma_s, \gamma_u\right)=
	\min\left(  \bar{\gamma}_{s}|h_s|^2,  \bar{\gamma_u}\abs{\sum_{i=1}^{N} \lambda_{s_i} \kappa_{s_i} }^2\right)
	.
\end{align}

\subsection*{Performance Metrics}
\subsubsection{Outage Probability}
The e2e outage probability of the considered system model \figurename{ \ref{SY2}} is given as
\begin{align} \label{Pout2}
	P_{out}&=\mathrm{Pr}(\min({\gamma_s},{\gamma_u})\leq{\gamma_{th}})=	F_{\gamma_{su}}(\gamma), \nonumber\\
	&=1-\left(1-F_{\gamma_s}(\gamma_{th})\right)(1-F_{\gamma_u}(\gamma_{th}))=F_{\gamma_s}(\gamma_{th})+F_{\gamma_u}(\gamma_{th})-F_{\gamma_s}(\gamma_{th})F_{\gamma_u}(\gamma_{th}).
\end{align} 
\begin{lemma} \label{l7}
	For sufficiently large number of IRS elements, the CDF of the e2e link is given as
	\begin{align} \label{CDF2}
		F_{\gamma_{su}}(\gamma)=1-Q_{\frac {1}{2}}\left ({\frac { {\mu_{\gamma_u} }}{\sigma_{\gamma_u} },\frac {\sqrt {\gamma}}{\sqrt {\bar {\gamma }}\sigma_{\gamma_u} }}\right) Q_{1}\left ({ \frac { \mu_{\gamma_s} } {\sigma_{\gamma_s} }, \frac {\gamma}{\sigma_{\gamma_s} } }\right), \quad  \gamma > 0
	\end{align}
\end{lemma}	
\begin{proof}
	Given in Appendix \ref{OP2}.  \qedhere
\end{proof}

\subsubsection{ASER Analysis}	
For ASER analysis,  \eqref{CDF2} can be re-written in series form \cite[eq. 18]{annamalai2009new} as
\begin{align}
	P_{out}&=1-A_3 \exp[-\Omega  \gamma_{th} ] \Gamma(M_2+k_2,\beta_{2}^{2}/2) \gamma_{th}^{n}, \label{CDF21}
\end{align}
where  $M_2=\frac{1}{2}$, $\alpha_{2}=\frac { \mu_{\gamma_u} } {\sigma_{\gamma_u}}$, $\beta_{2}=\frac {\sqrt {\gamma_{th}}} {\sqrt {\bar {\gamma_u }}\sigma_{\gamma_u} } $, and  $\Omega=\left({K+1 \over \gamma_{SR} }  \right)$.

\subsubsection*{Remark}
The above series is terminated at l=20 and verified in Mathematica. The series $k_2$ is terminated at 150 and verified by both Mathematica and Matlab.

\subsubsection*{Hexagonal QAM}
For system model \ref{SY2}, the generalized ASER expression of HQAM can be obtained in a similar manner as shown in  Appendix {\ref{H}}. Also, we apply the identities \cite[eq. (3.351.3), (7.522.9), (6.455.1), (9.14.1)]{gradshteyn2014table} to obtain the  generalized ASER expression of HQAM as
\begin{align}
{{P}_{s_2}}^{HQAM}&={1\over 2}\big( {{H}_{a}}-H_b \big) + \frac{{{H}_{b}}}{3} -\frac{{{H}_{b}}}{2} -  \frac{2{{H}_{b}}\alpha_{h}}{9\pi }  \frac{3}{2\alpha_{h}} {}_{2}{{F}_{1}}\Big( 1,1,\frac{3}{2},\frac{1}{2} \Big)
+\frac{{{H}_{b}}\alpha_{h} }{2\sqrt{3}\pi }  \frac{3}{2\alpha_{h}} 
\nonumber\\&\times
\left( {}_{2}{{F}_{1}}\Big( 1,1,{3 \over 2},\frac{1}{2} \Big)+{}_{2}{{F}_{1}}\Big( 1,1,\frac{3}{2},\frac{1}{4} \Big) \right)
+A_3 \bigg( \alpha^\nu  \bigg(  {\Gamma(\mu+\nu)\over \mu}\bigg( \frac{1}{2}\sqrt{\frac{\alpha_{h} }{2\pi }}
\nonumber\\&\times
\big( {{H}_{b}}-H_a \big) \Bbb{F}_2(\mu,\alpha,\beta_{1})  - \frac{{{H}_{b}}}{3}\sqrt{\frac{\alpha_{h} }{3\pi }} 
\Bbb{F}_2(\mu,\alpha,\beta_{2}) +
\frac{{{H}_{b}}}{2}\sqrt{\frac{\alpha_{h} }{6\pi }} \Bbb{F}_2(\mu,\alpha,\beta_{3})  \bigg) + \sum\limits_{i}
\nonumber\\&\times
\frac{ \Gamma(\mu_2+\nu)}{\mu_2 } \Bbb{F}_2(\mu_2,\alpha,\beta_{4})  \bigg( \frac{2{{H}_{b}}\alpha_{h}}{9\pi } {\left(\frac{\alpha_{h}}{3}\right)}^{i} -  \frac{{{H}_{b}}\alpha_{h} }{2\sqrt{3}\pi } 
\left\{ { \left(\frac{\alpha_{h}}{2} \right)}^{i} + { \left(\frac{\alpha_{h}}{6} \right)}^{i}  \right\} \bigg) \bigg)  \bigg),
\end{align}
where 	$\nu=\frac{1}{2}+k_1$; $\alpha=B_{\gamma_u} $, $\mu=n+\frac{1}{2}$, $\beta_1={{\frac{\alpha_{h}   }{2}}}+\Omega$, $\beta_2={{\frac{\alpha_{h}   }{3}}}+\Omega$, and $\beta_3={{\frac{\alpha_{h}   }{6}}}+\Omega$. Further, $\mu_2=n+i+1$ and $\beta_4={{\frac{2 \alpha_{h} }{3}}}+\Omega$.
\subsubsection*{Rectangular QAM}
For system model \ref{SY2}, the generalized ASER expression of RQAM can be obtained in a similar manner as shown in  Appendix {\ref{R}}. Also, we apply the identities \cite[eq. (3.351.3), (7.522.9), (6.455.1), (9.14.1)]{gradshteyn2014table} to obtain the  generalized ASER expression of RQAM as
\begin{align}
{{P}_{s_2}}^{RQAM}&= I_R +\frac{a_r b_r {R_1}{R_2}}{\pi}  \beta_3^{-1}  \bigg\{ {}_{1}{{F}_{1}}\left( 1,\frac{3}{2},{B_r \over   \beta_3  }   \right ) + {}_{1}{{F}_{1}}\left(1, 1,\frac{3}{2},{A_r \over  \beta_3  }  \right ) \bigg\}
\nonumber\\&
%
+A_3 \bigg( \alpha^\nu  \bigg(  {\Gamma(\mu+\nu)\over \mu}\bigg(  \frac{a_r{R_1}( {R_2}-1 )}{\sqrt{2\pi  }} \Bbb{F}_2(\mu,\alpha,\beta_{1})  + \frac{b_r {R_2}( {R_1}-1)}{\sqrt{2\pi}} \Bbb{F}_2(\mu_2,\alpha,\beta_{2}) \bigg)  
\nonumber\\&
-
\sum\limits_{i} \frac{a_r b_r {R_1}{R_2}}{\pi}  	\frac{ \Gamma(\mu_2+\nu)}{\mu_2}
\Bbb{F}_2(\mu_2,\alpha,\beta_{4})
\left( B_r^{i} + A_r^{i}  \right)  \bigg)  \bigg), 
\end{align}
where 	 $\nu=\frac{1}{2}+k_1$; $\alpha=\frac { 1} {2{\bar {\gamma_u }}\sigma_{\gamma_u}^2 }$, $\mu=n+\frac{1}{2}$, $\beta_1=A_r+\Omega$, $\beta_2=B_r+\Omega$, and $\beta_3=\frac{( {{a_r  }^{2}}+{{b_r }^{2}} ) }{2}$. Further, $\mu_2=n+i+1$, $\beta_4=\frac{( {{a_r  }^{2}}+{{b_r }^{2}} ) }{2}+\Omega$.
\subsubsection*{Cross QAM}
For system model \ref{SY2}, the generalized ASER expression of XQAM can be obtained in a similar manner as shown in  Appendix {\ref{X}}. Also, we apply the identities \cite[eq. (3.351.3), (7.522.9), (6.455.1), (9.14.1)]{gradshteyn2014table} to obtain the  generalized ASER expression of XQAM as

\begin{align}
P_{s_2}^X&=\frac{-1}{2}\mathbb{A}_X +\frac{2}{{M_{x}N_{x}}} 
+ \frac{16}{M_{x}N_{x}} \sum\limits_{l} \frac{l}{\pi \alpha_{x} } A_{x_3}^{-1} 
\left( {}_{1}{{F}_{1}}\Big( 1,\frac{3}{2},\frac{1}{A_{x_3} \alpha_{x}} \Big) + {}_{1}{{F}_{1}}\Big( 1,\frac{3}{2},\frac{4 l^2}{A_{x_3} \alpha_{x}} \Big) \right)	
\nonumber\\&
-2\frac{A_{x_2}}{\pi\alpha_{x}} \beta_4^{-1} \bigg( {}_{1}{{F}_{1}}\Big( 1,\frac{3}{2},\frac{A_{x_1}^2}{\beta_4 \alpha_{x}} \Big)
+{}_{1}{{F}_{1}}\Big( 1,\frac{3}{2},\frac{1}{\beta_4 \alpha_{x}} \Big) \bigg)
+\frac{k_x}{\pi\alpha_{x}} {\beta_5}^{-1}{}_{1}{{F}_{1}}\Big( 1,\frac{3}{2},\frac{1}{\beta_5 \alpha_{x}} \Big)   \bigg)
\nonumber\\&+
A_3 \bigg( \alpha^\nu  \bigg(  {\Gamma(\mu+\nu)\over \mu}\bigg( \frac{\mathbb{A}_X \Bbb{F}_2(\mu,\alpha,\beta_{1}) }{2\sqrt{\pi \alpha_{x}}}
-\frac{A_{x_2}}{\sqrt{\pi\alpha_{x}}} \Bbb{F}_2(\mu,\alpha,\beta_{2})  \bigg)  
+ \sum\limits_{n}  \frac{ \Gamma(\mu_2+\nu)}{\mu_2}  \bigg(   \frac{-16}{M_{x}N_{x}} 
\nonumber\\&\times
\sum\limits_{l}  \frac{l}{\pi \alpha_{x}} \Bbb{F}_2(\mu_2,\alpha,\beta_{3})  \left( { \left( \frac{1}{\alpha_{x}} \right) }^{n}  + {\left( \frac{4l^2}{\alpha_{x}}  \right)}^{n}  \right)  
-2\frac{A_{x_2}}{\pi\alpha_{x}} \Bbb{F}_2(\mu_2,\alpha,\beta_{4}) 	\left( { \left( \frac{1}{\alpha_{x}} \right) }^{n}  + {\left( \frac{A_{x_1}^2}{\alpha_{x}}  \right)}^{n}  \right)
\nonumber\\&
-\frac{k_x}{\pi\alpha_{x}} \Bbb{F}_2(\mu_2,\alpha,\beta_{5}) { \left( \frac{1}{\alpha_{x}} \right) }^{n}  \bigg) \bigg)   \bigg),
\end{align}
where $\nu=\frac{1}{2}+k_1$; $\alpha=\frac { 1} {2{\bar {\gamma_u }}\sigma_{\gamma_u}^2 }$, $\mu_1=n+\frac{1}{2}$,
$\beta_1=\frac{1}{\alpha_{x}}+\Omega $, and  $\beta_2=\frac{A_{x_1}^2}{\alpha_{x}}+\Omega $. Further, $\mu_2=n+i+1$, $\beta_3= A_{x_3}+\Omega$, $\beta_4=\frac{1+A_{x_1}^2}{\alpha_{x}}+\Omega $, and $\beta_5= \frac{2}{\alpha_{x}}+\Omega$.

\subsubsection*{Ergodic rate analysis}
The ergodic rate for system model system model \ref{SY2} can be obtained  by substituting the e2e SNR of system model system model \ref{SY2} in  \eqref{Er}.
%
\begin{lemma}\label{l8}
The generalized ergodic rate expressio of the system model \ref{SY2} is given as
\begin{align}\label{ERG2}
	C_{R_ 2} &= {A_3\over{2\ln2} } \sum_{j=0}^{\infty} {(-\Omega)^j \over j!} 
	{1^{n+j} \over \Gamma(1)}G_{1+1,2+1}^{2+1,0+1} \left( \begin{matrix}
		\beta
	\end{matrix} \bigg| \begin{matrix}
		1-\rho, a_1,...,a_p \\ \sigma-\rho, b_1,...,b_q
	\end{matrix} \right), 
\end{align}
\end{lemma}
where 
$A_3= e^{-\alpha_{2}^{2}/2}\sum_{k_2=0}^{\infty} \sum_{l=0}^{\infty}\sum_{n=0}^{l}\frac{K_{i}^{l}\Omega_{i}^{n}}{(l)!n!} {1 \over k_2!}\left({\alpha_{2}^{2} \over 2}\right)^{k_2}{ \exp[-K]  \over \Gamma(M_{2}+k_{2})}$. Further, $\rho= n+j+1$,   $\beta_1=1$, $\sigma=1$, $\alpha=\beta$, $m=2, n=0, p=1$, and $ q=2$.
\begin{proof}
Proof is given in Appendix \ref{ER2} \qedhere
\end{proof}
\newsavebox\correlated
\begin{lrbox}{\correlated}
{\small{
		$ \begin{aligned}
			&e^{-\alpha_{2}^{2}/2}\sum_{k_2=0}^{\infty} \sum_{l=0}^{\infty}\sum_{n=0}^{l}\frac{K_{i}^{l}\Omega_{i}^{n}}{(l)!n!} {1 \over k_2!}\left({\alpha_{2}^{2} \over 2}\right)^{k_2}{ \exp[-K]  \over \Gamma(M_{2}+k_{2})}
		\end{aligned} $ 
}}
\end{lrbox} 
	\newsavebox\mc
	\begin{lrbox}{\mc}
{\small{
		$ \begin{aligned}
			\sum\limits_{y=0}^{\infty }{\frac{{{( 1 )}_{y}}}{{{( \frac{3}{2} )}_{y}}y!}}; \quad
			\frac{1 }{{{( {{b}_{1}}+{{c }_{1}} )}^{{a_1}+\vartheta}}}{}_{2}{{F}_{1}}\big( 1,{{a_1}+{\vartheta}},{a_1+1},\frac{{c_1}}{{b_1}+{c_1}}\big).
		\end{aligned}  $ 
}}
\end{lrbox} 

\section{Numerical Analysis}
In this section, the numerical results are presented and the accuracy of the analytical expressions is evaluated with the Monte-Carlo simulations. Unless otherwise stated, the considered simulation parameters are given in  \tablename{ \ref{P}}. In figures, `Ana', Asym., and `Sim', indicate the `analytical', 'asymptotic' and `simulation results', respectively. Further 'SM-1' and 'SM-2' indicate the considered system models 1, 2 (as shown in \figurename{ \ref{SY1}}  and    \figurename{ \ref{SY2}}), respectively. In figures, Analytical results are represented by solid lines whereas the simulation results are represented by markers. 

\vspace{-1.5em}

\begin{table*}
\caption{Functional representation in Arithmetic expressions}
\centering
\begin{threeparttable}		
	\begin{tabular}{cc@{\qquad}}
		\toprule
		\textbf{Function}  & \textbf{Representation} \\ \bottomrule  
		\makecell{$A_3$ \\} &
		\usebox{\correlated}   \\ \cmidrule(l r){1-2}
		%
		%
		$\sum\limits_{y=0}^{}$;\hspace{1em} $\Bbb{F}_2(a_{1},b_{1},c_{1})$ & \usebox{\mc} \\ 
		\midrule\midrule	
	\end{tabular}	 
\end{threeparttable}
\end{table*}
\begin{figure}[H]
\centering
\includegraphics[width=3.25in]{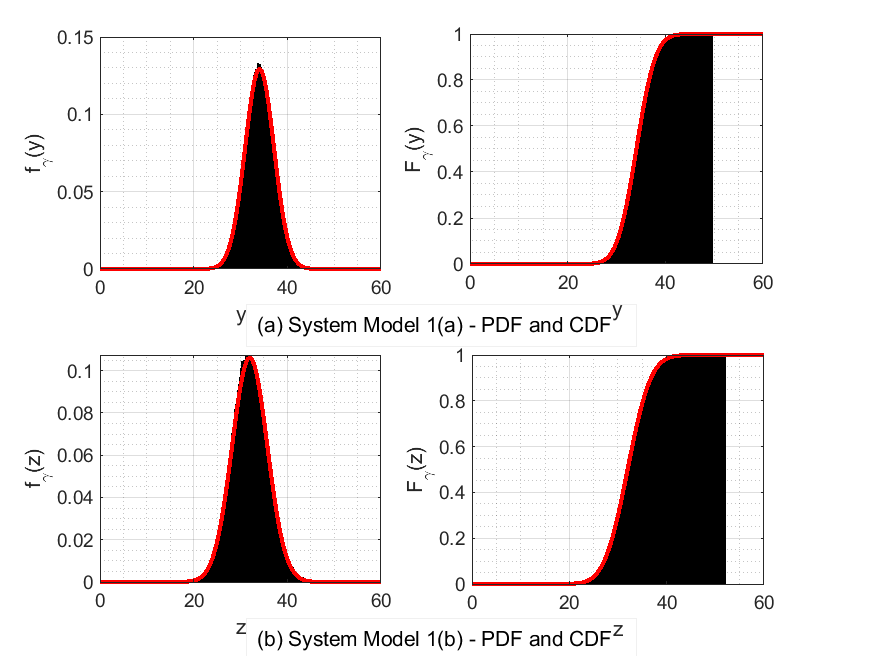}\quad
\caption{PDFs and CDFs of IRS links of system models 1, 2 (as shown in \figurename{ \ref{SY1}}  and    \figurename{ \ref{SY2}}).\vspace{-0.5em}}
\label{CPDF}
\end{figure}
\begin{table}[]
\caption{Simulation Parameters}
\label{P}
\begin{tabular}{|l|l|l|ll}
	\cline{1-3}
	Parameters & System model 1 &System model 2 &  &  \\  \cline{1-3}
	carrier frequency $f_\text{c}$ &    $5 $ GHz                    &  $5 $ GHz                           &  &  \\ 
	Light velocity $\text{c}$&     $3 \times 10^8 $ m/s                  &    $3 \times 10^8 $ m/s                    &  &  \\ 
	Height of satellite &         $35,786$ Km               &  $35,786$ Km                       &  &  \\ 
	Height of  HAP &        $20$ Km               &     $20$ Km                    &  &  \\ 
	$G_\text{Max}$&             $56$ dBi     &         $56$ dBi                &  &  \\ 
	$\{\phi_\text{SH}, \phi_\text{3dB}\}$&                 $\{0.4^{\circ}, 0.8^{\circ} \}$  &            $\{0.4^{\circ}, 0.8^{\circ} \}$            &  &  \\ 
	$G_\text{H}$ &                     $\text{N}_\text{IRS}$   &    105 dB  \cite{3gpp,xing2021high}                  &  &  \\ 
	user gain 	&                   $G_{Rx} =2$    &                 $Gr_\text{U} =  2$    &  &  \\ 
	$L_\text{Rain}$  &   $0.01$ dB/Km                       &     $0.01$ dB/Km                     &  &  \\ 
	$	L_\text{Atm.}$	&    $5.4\times10^{-3}$ dB/Km                   &       $5.4\times10^{-3}$ dB/Km                  &  &  \\ 
	$L_\text{oth}$	&     $2$ dB                   &        $2$ dB                       &  &  \\ 
	$\{h_t, h_r\}$	&      -                &             \{50, 5\} m              &  &  \\ 
	$d_{\text{I}_\text{N}\text{U}}$		&                      &         300m                   &  &  \\ 
	Heavy shadowing (HS)		&    $b_0=0.063$   $m_0=1$   $\Omega_0=0.0007$                  &   $b_0=0.063$   $m_0=1$   $\Omega_0=0.0007$                                &  &  \\ 	
	Average shadowing (AS)				&       $b_0=0.251$   $m_0=5$   $\Omega_0=0.279$	                &     $b_0=0.251$   $m_0=5$   $\Omega_0=0.279$	                          &  &  \\ 
	Light shadowing	(LS)			&    $b_0=0.158$   $m_0=19$   $\Omega_0=1.29$                   &     $b_0=0.158$   $m_0=19$   $\Omega_0=1.29$                             &  &  \\ 	\cline{1-3}
\end{tabular}
\end{table}
\vspace{-2em}

\begin{figure*}[htp]
\centering
\subfigure[OP  vs transmit power of SM-1 and SM-2 for 30 IRS elements.\label{OP_1}]{\includegraphics[width=2.85in]{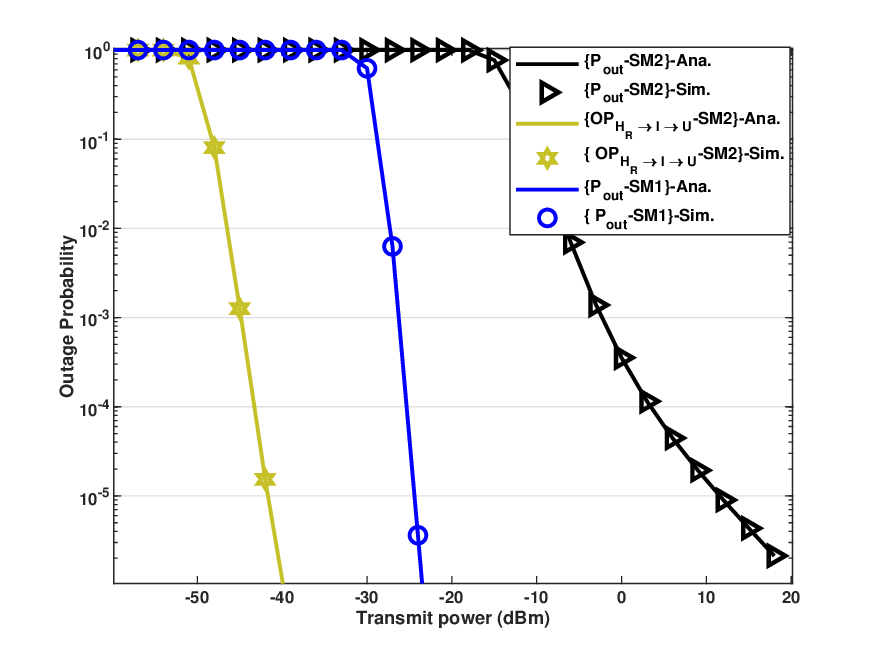}}\quad
\subfigure[OP  vs transmit power of SM-1 and SM-2 for various HAP gains.\label{OP_2}]{\includegraphics[width=2.85in]{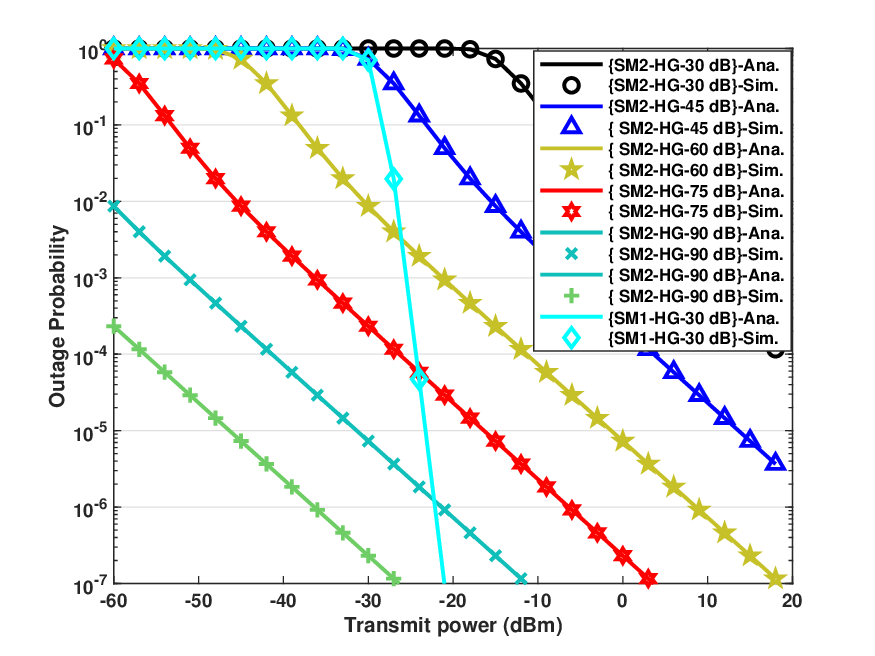}}\quad
\caption{Outage probability versus transmit power of SM-1 and SM-2.}
\label{OP_12}
\end{figure*} 

In \figurename{ \ref{CPDF}}, the PDFs and CDFs correspond to the IRS links of system models 1  and  2, respectively are presented to verify the central limit theorem (CLT). In \figurename{ \ref{CPDF}\textcolor{red}{(a)}},  CLT is verified through the histogram plot for both PDF and CDF correspond to   $\text{S} \rightarrow \text{H}_\text{I} \rightarrow \text{U}$ link with Rician and SR fading.  In \figurename{ \ref{CPDF}\textcolor{red}{(b)}},  CLT is verified through the histogram plot for both PDF and CDF corresponding to   $\text{H}_\text{R} \rightarrow \text{I} \rightarrow \text{U} $ link with SR and Nakagami-$m$ fading.

\begin{figure*}[htp]
\centering
\subfigure[OP performance of SM-1 for various IRS elements.\label{OP_3}] {\includegraphics[width=2.85in]{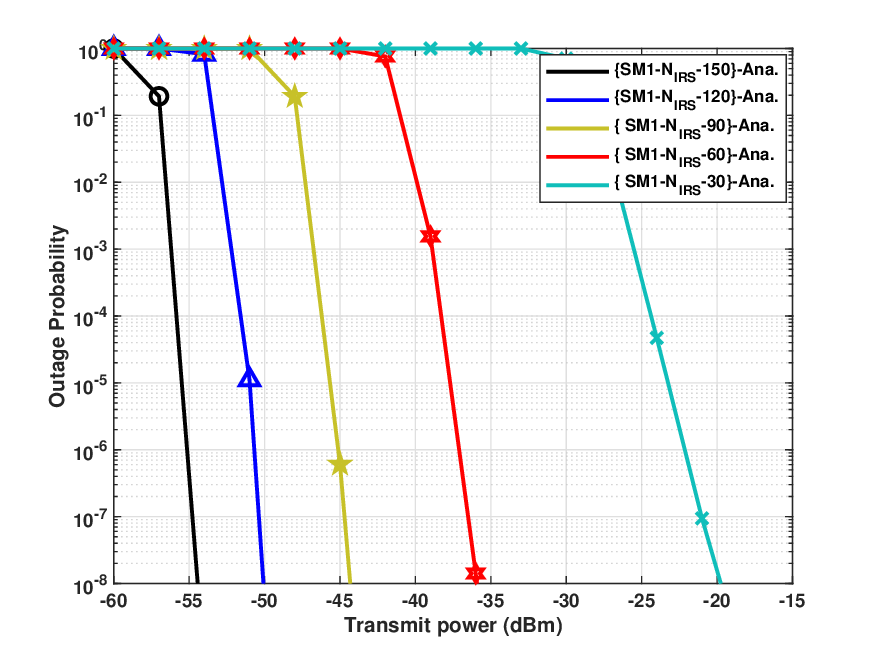}}\quad
\subfigure[OP performance of SM-1 for various various Rician parameters under different shadowing conditions. \label{OP_4}]{\includegraphics[width=2.85in]{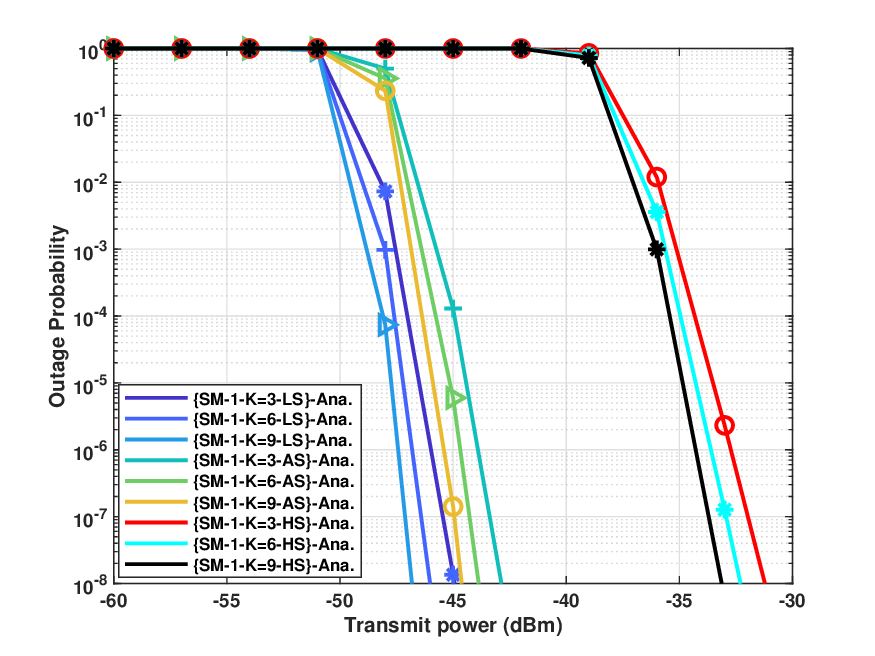}}\quad
\caption{Impact of IRS elements and the Rician factor over OP  performance of SM-1.\vspace{-1em}}
\label{OP_22}
\end{figure*}

The OP curves are presented in \figurename{ \ref{OP_12}} for both system models under HS fading  with $30$ IRS elements. In \figurename{ \ref{OP_1}}, for SM-1, HAP gain is considered to be IRS elements gain whereas in  SM-2, HAP gain of $30$ dB is considered. The overall system performance of SM-2 is dominated by the $\text{S} \rightarrow \text{H}_\text{R} $ link performance  than the $\text{H}_\text{R} \rightarrow \text{I} \rightarrow \text{U} $ link. Thus, the system with the  aerial assisted IRS node (SM-1) outperforms the system with the terrestrial assisted IRS node (SM-2). In \figurename{ \ref{OP_2}}, SM-1 OP results are compared with SM-2 OP results for various HAP gains.To ensure a fair comparison, we consider $30$ IRS elements and set $K$ to $5$ in the SM-1 scenario. IRS elements gain constitutes the HAP gain for SM-1. Results illustrate that with the increase in HAP gain from $30$ dB to $105$ dBm \cite{3gpp}, OP performance of SM-2 improves over the SM-1 OP performance. Finally, simulation results match well with the analytical results.

Results in \figurename{\ref{OP_22}}, demonstrate the impact of IRS elements and the Rician parameter over the OP of the SM-1. In \figurename{ \ref{OP_3}}, OP versus transmit power is presented for SM-1 for different IRS elements. For an OP of $10^{-8}$, system with $150$ IRS elements provides a  transmit power gain of $\approx 35$ dBm  and $\approx 4$ dBm as compared to the system with $30$ and $120$ IRS elements, respectively. The increase in transmit power gain reduces with the increase in IRS elements. In \figurename{ \ref{OP_4}}, OP performance of SM-1 is presented with respect to the transmit power for different Rician fading paramters and shadowing effects. For an OP of $10^{-8}$, system with K=3 has a transmit power gain of  $0.75$ dBm  and $\approx 1.2$ dBm  over $K=6$ and $K=9$, respectively (irrespective to the shadowing conditions). Further, for an OP of $10^{-8}$, LS has a transmit power gain of $\approx 12$ dBm and $\approx 14$ dBm over AS and HS conditions.

\begin{figure*}[htp]
\centering
\subfigure[Impact of IRS elements.\label{ER_1a}] {\includegraphics[width=2.85in]{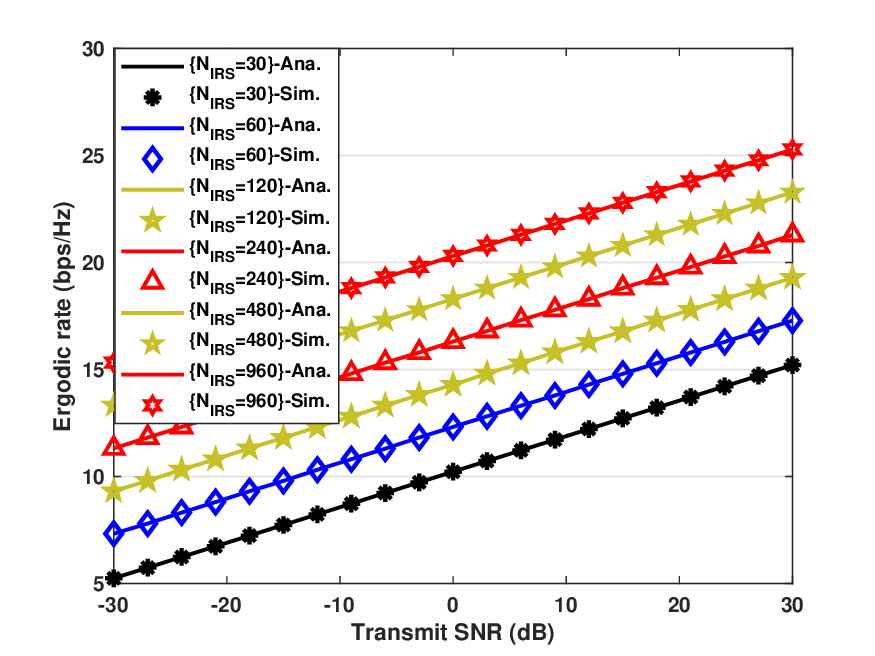}}\quad
\subfigure[Impact of shadowing. \label{ER_1b}]{\includegraphics[width=2.85in]{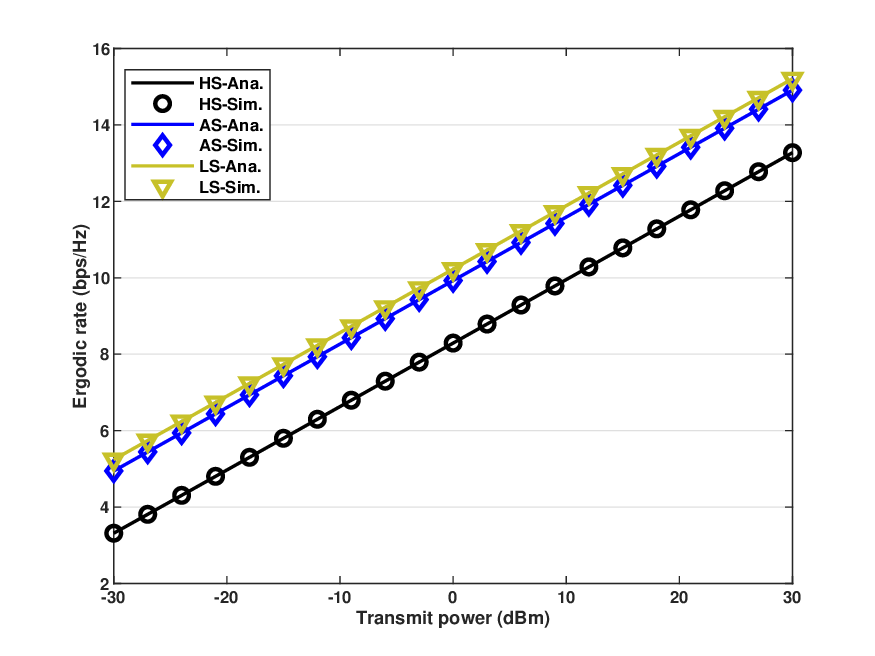}}\quad
\caption{Ergodic rate versus transmit power  of SM-1.\vspace{-1.5em}}
\label{ER_1}
\end{figure*}

Ergodic rate analysis is shown in \figurename{ \ref{ER_1}} and illustrated   the effect of fading conditions and the IRS elements for SM-1 with $K=5$. In \figurename{ \ref{ER_1}\textcolor{red}{(a)}}, the curves depicts the affect  of IRS elements under LS over ergodic rate. With an increase in IRS elements from $30$ to $960$, the ergodic rate increases from $15$ bps/Hz to $25$ bps/Hz. With every two-folds increase in IRS elements, the ergodic rate increases by $\approx 2$ bps/Hz.   The effects of shadowing are shown in \figurename{ \ref{ER_1}\textcolor{red}{(b)}}.  For $30$ dBm transmit power, system under LS has ergodic rate gain of $\approx 1.93$ bps/Hz over the system under HS. Finally, the derived  simulation  results matches closely with the analytical results.

\begin{figure*}[htp]
\centering
\subfigure[Ergodic rate of SM-1 and SM-2 for 30 IRS elements.\label{ER_2}] {\includegraphics[width=2.85in]{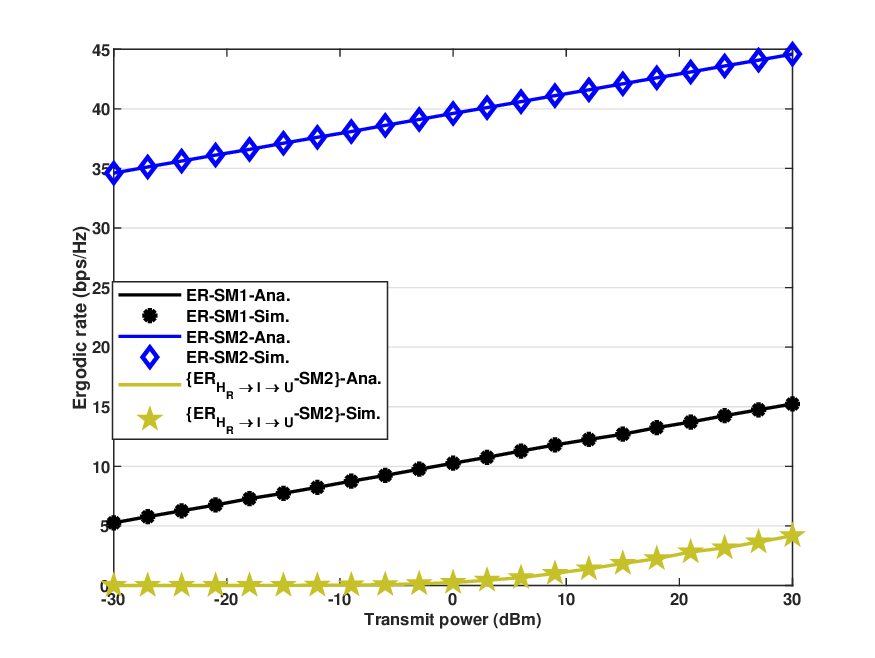}}\quad
\subfigure[Ergodic rate of SM-2 for various Rician parameters values and HAP gains. \label{ER_3}]{\includegraphics[width=2.85in]{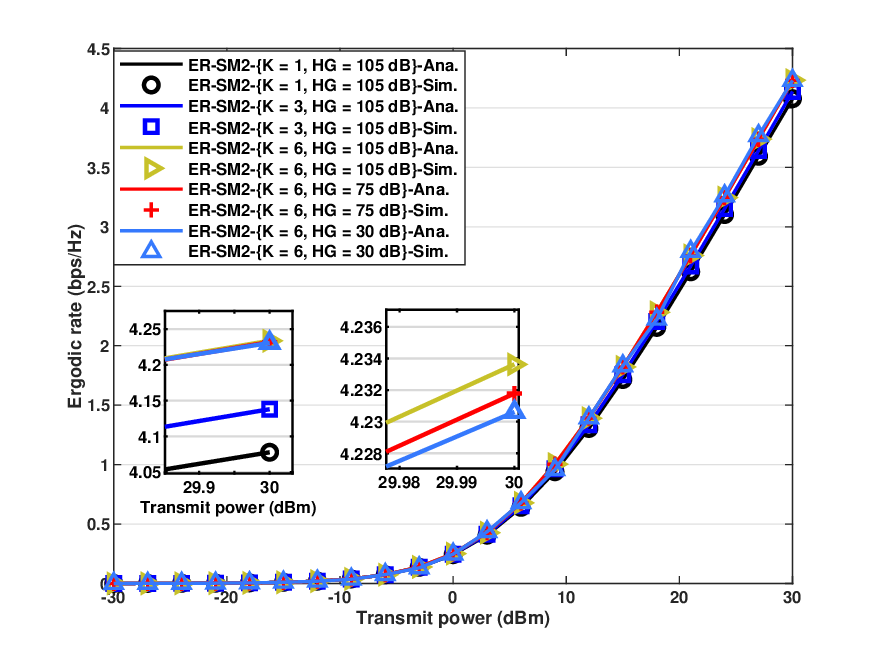}}\quad
\caption{Ergodic rate versus transmit power.\vspace{-1.5em}}
\label{ER_SM2}
\end{figure*}
In \figurename{ \ref{ER_SM2}}, ergodic rate analysis of the system model 2 is presented with respect to the transmit power  under LS. . The comparison between the ergodic rates of   SM-1 and SM-2 are depicted in \figurename{ \ref{ER_2}}. For SM-1, $30$ IRS elements are considered, while SM-2 utilizes a HAP gain of $105$ dBm. The results highlight that the overall ergodic rate achieved in SM-2 is dominated by the $\text{S} \rightarrow \text{H}_\text{R}$ link, exhibiting a rate of $4.17$ bps/Hz, surpassing the ergodic rate of $44.59$ bps/Hz for the $\text{H}_\text{R} \rightarrow \text{I} \rightarrow \text{U} $ link at $30$ dBm transmit power. Thus, the aerial IRS node based SM-1 outperforms  the terrestrial IRS node based SM-2 with a ergodic rate of $\approx 15$ bps/Hz.  In \figurename{ \ref{ER_3}}, ergodic rates of SM-2 are presented for Rician parameters $K=\{1,3,6\}$ for a HAP gain of 105 dBm. Additionally, ergodic rates for $K=6$ with HAP gains ${105, 75, 30}$ dBm are shown to realize the effects of both $K$ parameters and HAP gain on the ergodic rates. The results indicate that, for a transmit power of $30$ dBm and a HAP gain of $105$ dBm, the system with $K=6$ achieves an ergodic rate gain of $0.15$ bps/Hz over the system with $K=1$. This marginal improvement signifies the changes in the HAP gain have limited impact on the overall performance. For $K=6$, the system with a HAP gain of $105$ dBm has an ergodic rate gain of  $\approx 0.0025$ bps/Hz over a system with a HAP gain of $30$ dBm. This observation is attributed to the fact that the overall diversity gain of SM-2 is dominated by the diversity gain of the $\text{S} \rightarrow \text{H}_\text{R}$ link, while in SM-1, the ergodic rates attained are directly proportional to the diversity gain of IRS elements.

\begin{figure}
\centering
\begin{minipage}{.47\textwidth}
	\centering
	\includegraphics[width=2.9in]{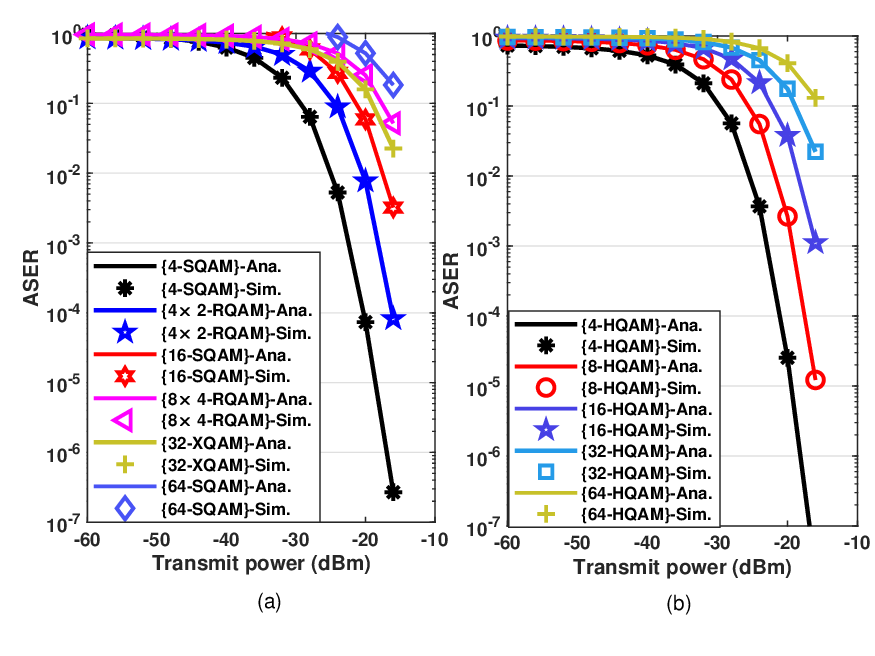}\quad
	\caption{ASER  vs transmit power of SM-1 
		for various constellations.\vspace{-1.5em}}
	\label{ASER1}
\end{minipage}%
\quad
\begin{minipage}{.47\textwidth}
	\centering
	\includegraphics[width=2.9in]{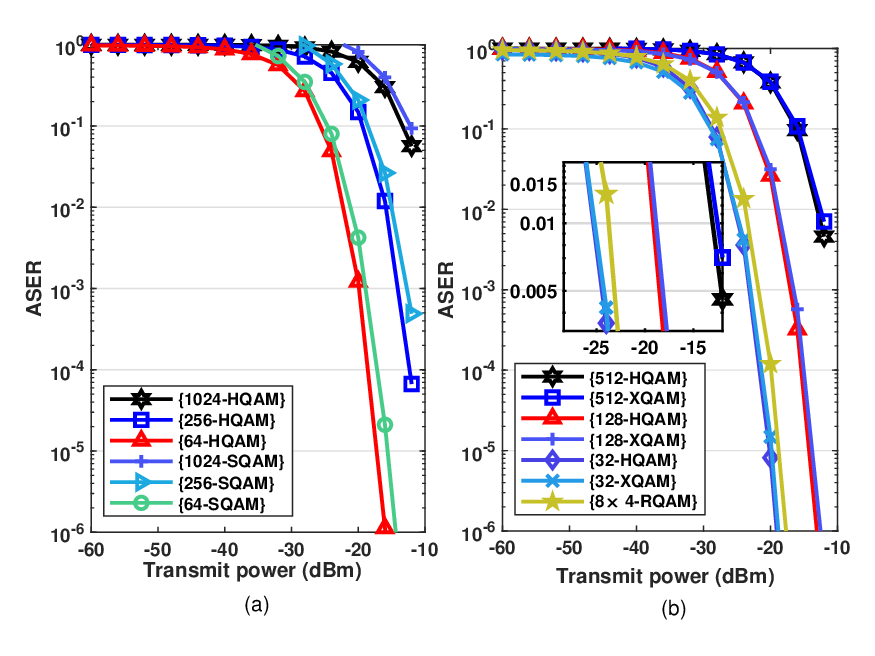}\quad
	\caption{ASER  vs transmit power of SM-1 for higher order constellation sizes.\vspace{-1.5em}}
	\label{ASER2}
\end{minipage}
\end{figure}

ASER results  for various QAM schemes are demonstrated in \figurename{  \ref{ASER1}} for SM-1 under HS scenario for $K= 3$. In \figurename{ \ref{ASER1}}\textcolor{red}{(a)} ASER analysis of  SQAM, RQAM, and XQAM are shown whereas in \figurename{ \ref{ASER1}}\textcolor{red}{(b)} HQAM results are plotted for constellation points $4$, $8$, $16$, $32$, and $64$.  Monte-Carlo simulations closely align with the derived analytical results, demonstrating good agreement. For an ASER of $10^{-2}$, $32$-XQAM has a transmit power gain of $\approx 2$ dBm over  $8 \times 4$ RQAM. 

In \figurename{ \ref{ASER2}}, even and odd higher order constellations ASER results  are shown for various QAM schemes under AS scenario  with $30$ IRS elements and $K = 3$. In \figurename{ \ref{ASER2}}\textcolor{red}{(a)}, the even constellation points $64$, $256$, and $1024$ are presented.  The results clearly demonstrate that as the constellation sizes increase from $64$, $256$, and $1024$, HQAM outperforms the SQAM with a transmit power gain of $\approx 1$ dBm. This improvement is attributed to the  optimum 2 dimensional hexagonal lattice of HQAM with minimum peak and average power than the  SQAM. In \figurename{ \ref{ASER2}}\textcolor{red}{(b)}, the ASER analysis of  RQAM, XQAM, and  HQAM is presented with odd constellation points ($32$, $128$, and $512$). It is observed that for odd bits transmission also, HQAM outperforms both RQAM and XQAM. For an ASER of $10^{-4}$, $32$ HQAM has transmit power gain of $\approx 0.25$ dBm over the XQAM and $\approx 2$ dBm over $8 \times 4$ RQAM. Additionally, for an ASER of $10^{-2}$, HQAM has transmit power gain of $\approx 0.2$ dBm for $128$ and $512$ constellations sizes, respectively. These findings indicates the superiority of HQAM in achieving improved performance across different constellation sizes.
\begin{figure*}[htp]
\centering
\subfigure[ASER analysis of HQAM for SM-1 and SM-2.\label{HQAM}] {\includegraphics[width=2.85in]{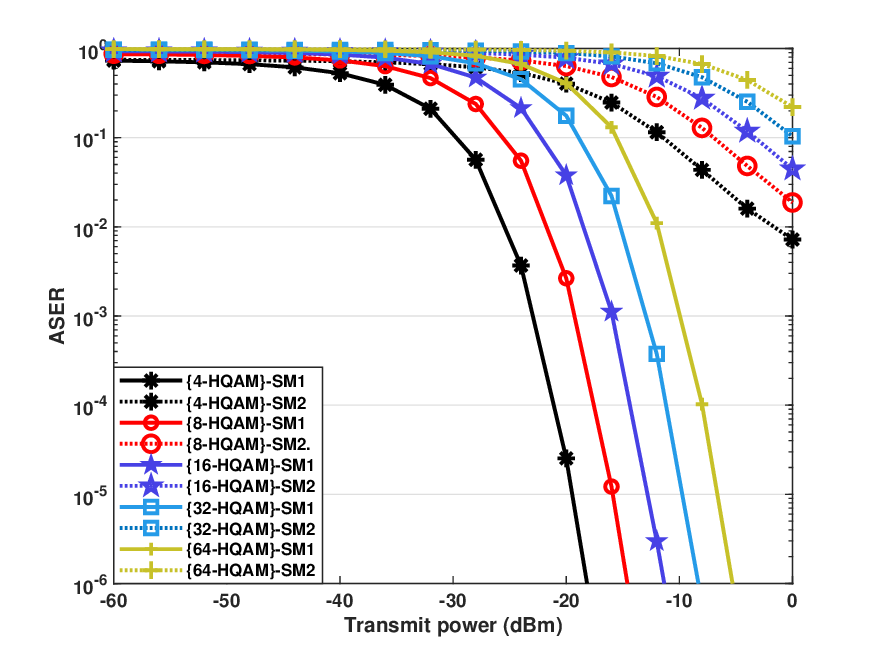}}\quad
\subfigure[ASER analysis of 16 and 32 constellation points \label{ASER3}]{\includegraphics[width=2.85in]{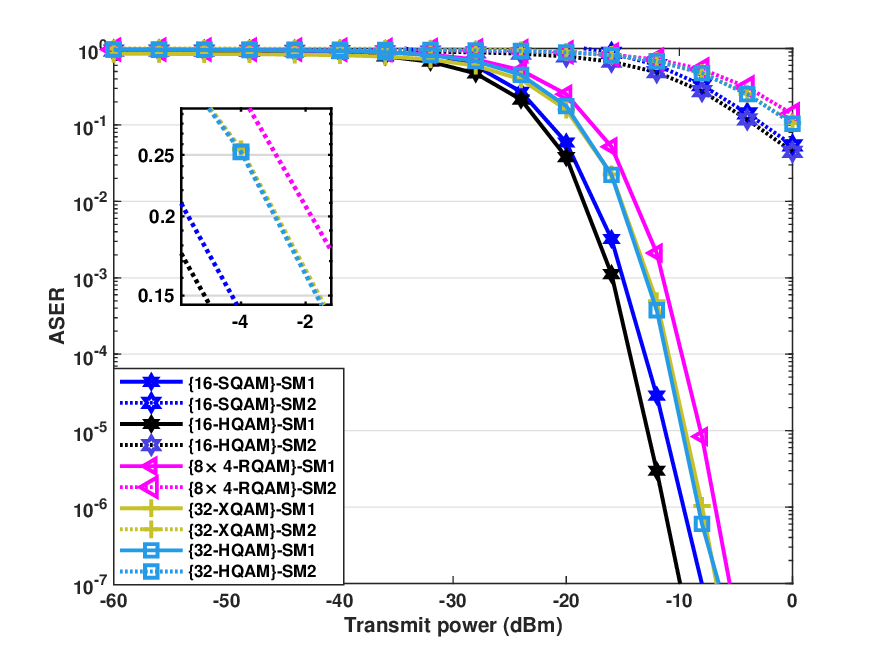}}\quad
\caption{ASER analysis of various modulation schemes.\vspace{-1.5em}}
\label{ASER_12}
\end{figure*}

In \figurename{ \ref{ASER_12}}, comparative ASER analysis results for various QAM schemes between SM-1 and SM-2  with 30 IRS elements under HS scenario are presented. In \figurename{ \ref{HQAM}},  ASER analysis of HQAM for both SM-1 and SM-2 is presented for constellation points $4$, $8$, $16$, $32$, and $64$. For an ASER of $10^{-1}$, for all the constellation points,  SM-1 consistently exhibits an average transmit power gain of $\approx 18$ dBm over SM-2. The overall system performance in SM-1 is significantly boosted with the IRS elements.  On the other hand, in SM-2, the system's performance is influenced by the $\text{S} \rightarrow \text{H}_\text{R}$ link, impacting its overall efficiency. In \figurename{ \ref{ASER3}}, ASER analysis of HQAM, SQAM, RQAM, and XQAM is represented for $16$ and $32$ constellation points for both SM-1 and SM-2. For an ASER of $10^{-3}$, 16-HQAM has a transmit power gain of $\approx 0.85$ dBm over 16-SQAM, while  for 32- constellation points, HQAM achieves a gain of $\approx 0.3$ dBm and $\approx 1.5$ dBm over the XQAM and RQAM schemes, respectively. For SM-2, an ASER of $10^{-1}$, $16$-HQAM  exhibits a transmit power gain of $\approx 0.6$ dBm over the $16$-SQAM. The performance gain of HQAM over other QAM schemes is due to its low peak and average energies with optimum 2 dimensional hexagonal constellation.

\section{Conclusion}
In this study, performance analysis of satellite-terrestrial networks with aerial and terrestrial IRS nodes over diverse fading channels, including shadowed Rician, Rician, and Nakagami-$m$ fading channels is presented. System performance is examined through the closed-form expressions of outage probability, ergodic rate, and  average symbol error rate for the higher-order modulation schemes. Practical antenna gains, path losses, and various link fading scenarios are taken into account to characterize the satellite-terrestrial links accurately. The analysis reveals that SM-1, operating under heavy shadowing conditions, outperforms SM-2, which operates under lighter shadowing. Aerial IRS node based system exhibits superior performance compared to the terrestrial IRS node based system. The overall ergodic rate attained in an aerial IRS node system is proportional to the IRS elements,  whereas in the terrestrial IRS node-based system, the overall ergodic rate is influenced by the diversity provided by the satellite-HAP link. By considering more HAP gain in SM-2, the overall system performance can be enhanced compared to SM-1. Additionally, in SM-1, a notable observation is that with every increase in IRS elements by two-folds, the ergodic rate increases by $\approx 2$ bps/Hz. It is also observed that for both even and odd bits transmission, HQAM outperforms the other QAM schemes.


\appendices
\section{Proof of Lemma \ref{l1}}
\begin{proof}\label{OP1}
In \eqref{a1}, it is observed that $\gamma$ is the sum of products of $\alpha_i$ and  $\beta_i$ which are independent Rician and shadowed Rician fading channels, respectively. To facilitate the mathematical tractability the   PDF and cumulative distribution function (CDF) of $\gamma$ are approximated tightly.
\subsection*{Statistical characterization of the optimal received SNR}
By invoking central limit theorem, in \eqref{a1}, the PDF and CDF of $\gamma$ are approximated  with $R=\bar{\gamma}Z^2$, where in  $Z=\sum_{i=0}^{N}\tilde{Z}$
Let $\tilde{Z}=XY$ where $X$ and $Y$ are independent random variables. Mean and variance of $\tilde{Z}$ are given as:
\begin{align}
	\mathrm{E}(\tilde{Z})=\mathrm{E}(X)\mathrm{E}(Y)&= \left(4\sigma^2_R b_0\right)^{1/2} \Gamma \left( {{3}\over {2}} \right) B_0^m {}_1{F_1}\left({ -\frac{1}{2},1;-K} \right) {}_2F_1 \left( {{3}\over {2}},\, m,\, 1,\, B_1 \right) ,
\end{align}
\begin{align}\label{VAR}
	\mathrm{Var}(\tilde{Z})&=\frac{2(2b_0) \bar{\gamma}_{i}^2}{(1+K)^2} B_0^m   {}_1{F_1}\left({ -2,1;-K} \right)
	{}_2F_1 \left( 2,\, m,\, 1,\, B_1 \right)-\frac{(2b_0)^{1/2} \bar{\gamma}_{i} }{(1+K)}\Gamma \left( {{3}\over {2}} \right)B_0^m
	\nonumber\\&\times
	{}_1{F_1}\left({ -1,1;-K} \right)
	{}_2F_1 \left( {{3}\over {2}},\, m,\, 1,\, B_1 \right),	
\end{align}
where $B_0 = \left( {{2b_0 m_h}\over {2b_0 m_h + \Omega}} \right) $ and $B_1={{\Omega }\over {2b_0 m_h + \Omega }}$. Mean and variance of $Z$  are given as $\mu_Z=\mathrm{E}(Z)=\sum_{i=1}^{N}\mathrm{E}(\tilde{Z}_i)$ and $\sigma_Z^2=\mathrm{Var}({Z})=\sum_{i=1}^{N}\mathrm{Var}(\tilde{Z}_i)$, respectively.
In practice, the meta-surface IRS elements  are of conformal geometry, light weight, low cost, and size. Hence, it is practically possible to use large number of reflecting surfaces.  Thus, for sufficiently large number of reflecting meta-surfaces, and as per central limit theorem (CLT) $Z^2$ follows a non-central chi-square random variable with one degree of freedom with mean $\mu_Z=N\mu_{\tilde{Z}}$ and variance $\sigma^2_Z=N\sigma^2_{\tilde{Z}}$. 
The PDF of $\gamma$ is given as 
\begin{align} \label{PDF1a}
	f_{\gamma }\left ({\gamma }\right)=\frac {1}{2\sigma_\gamma ^{2}\bar {\gamma }}\left ({\frac {\gamma }{\bar {\gamma }{\mu}_\gamma^2 }}\right)^{-\frac {1}{4}}{\mathrm {exp}}\left ({-\frac {\gamma +{\mu}_\gamma^2 \bar {\gamma }}{2\bar {\gamma }\sigma_\gamma ^{2}}}\right)\textit I_{-\frac {1}{2}}\left ({\frac {{\mu_\gamma}\sqrt {\gamma  }}{\sqrt {\bar {\gamma }}\sigma_\gamma ^{2}}}\right), \quad \gamma > 0
\end{align}
where $\mu_\gamma= \mu_Z$ and $\sigma_\gamma=\sigma_Z$. The CDF is given as \cite[eq. (2.3-35)]{proakis1995digital}
\begin{align}
	F_\gamma(\gamma)=1-Q_{\frac {1}{2}}\left ({\frac { {\mu_\gamma }}{\sigma_\gamma },\frac {\sqrt {\gamma}}{\sqrt {\bar {\gamma }}\sigma_\gamma }}\right), \quad  \gamma > 0
\end{align}	
\end{proof}
\vspace{-2em}
\section{Proof of Lemma  \ref{l3}}

\begin{proof} \label{H}
To derive the closed-form ASER expression for HQAM, the first order derivative of the conditional SEP  (\ref{HSEP}) is obtained by using the identities $Q(x)=\frac{1}{2}\Big[1- \text{erf}\big(\frac{x}{\sqrt2}\big)\Big]$ and\cite[Eq. (7.1.21)]{abramowitz1964handbook} and differentiating it w.r.t $\gamma$ as
\begin{align}\label{HFO}
	&{{P}_{s}}^{'}\big( e|\gamma  \big)=\frac{1}{2}\sqrt{\frac{\alpha_{h} }{2\pi }}\big( {{H}_{b}}-H_a \big){{\gamma }^{-\frac{1}{2}}}{{e}^{-\frac{\alpha_{h} \gamma }{2}}}-\frac{{{H}_{b}}}{3}\sqrt{\frac{\alpha_{h} }{3\pi }}{{\gamma }^{-\frac{1}{2}}}{{e}^{-\frac{\alpha_{h} \gamma }{3}}}+\frac{{{H}_{b}}}{2}\sqrt{\frac{\alpha_{h} }{6\pi }}{{\gamma }^{-\frac{1}{2}}}{{e}^{-\frac{\alpha_{h} \gamma }{6}}} +\frac{2{{H}_{b}}\alpha_{h}}{9\pi }\nonumber\\ 
	&\times{{e}^{-\frac{2\alpha_{h} \gamma }{3}}}{}_{1}{{F}_{1}}\Big( 1,\frac{3}{2},\frac{\alpha_{h}}{3}\gamma \Big) -\frac{{{H}_{b}}\alpha_{h} }{2\sqrt{3}\pi }{{e}^{-\frac{2\alpha_{h} \gamma }{3}}}\bigg\{{}_{1}{{F}_{1}}\Big( 1,\frac{3}{2},\frac{\alpha_{h}}{2}\gamma \Big)+{}_{1}{{F}_{1}}\Big( 1,\frac{3}{2},\frac{\alpha_{h} }{6}\gamma  \Big) \bigg\}. 
\end{align}
To derive the closed-form ASER expression for HQAM, the \eqref{CDF1} can be rewritten by taking the approximation of Marcum-Q function  as given in  \cite[eq. 18]{annamalai2009new} as
\begin{align}\label{CDF12}
	F_{\gamma_{e2e}}(\gamma)&=1- \sum_{k_1=0}^{\infty} e^{-\alpha_{1}^{2}/2}{1 \over k_1!}\left(\alpha_{1}^{2}\over2\right)^{k_1}{\Gamma(\frac{1}{2}+k_1,B_{\gamma_u} \gamma ) \over \Gamma(\frac{1}{2}+k_{1})}.
\end{align}
where   $B_{\gamma_u} = \left( \frac { 1} {2{\bar {\gamma_u }}\sigma_{\gamma_u}^2} \right)$. On substituting $P_s^{'}(e|\gamma)$ and $F_{\gamma_{e2e}}(\gamma)$ from (\ref{HFO}) and (\ref{CDF12}),  respectively into (\ref{SEP}), we get
\begin{align}
	{{P}_{s}}^{HQAM}=&-\int_{0}^{\infty}P'_{s}(e|\gamma)F_{\gamma_{e2e}}(\gamma) d\gamma,  
	\nonumber\\=&
	-\int_{0}^{\infty}P'_{s}(e|\gamma)d\gamma	
	+
	A_2\int_{0}^{\infty}P'_{s}(e|\gamma) 
	\Gamma\left(\frac{1}{2}+k_1,B_{\gamma_u} \gamma \right)   d\gamma
	= -P_{H_{1}} + P_{H_{2}}, \label{eq:HQ}
\end{align} 
$P_{H_{1}}$ can be resolved by using the identities \cite[eq. 3.351.3,  eq. 7.522.9]{gradshteyn2014table} and is given as
\begin{align}\label{Ph1}
	P_{H_{1}}&={ -{{H}_{a}}\over 2}+\frac{2{{H}_{b}}}{3} +  \frac{{{H}_{b}}}{3\pi } {}_{2}{{F}_{1}}\Big( 1,1,\frac{3}{2},\frac{1}{2} \Big)	+\frac{3{{H}_{b}} }{4\sqrt{3}\pi } 
	\left( {}_{2}{{F}_{1}}\Big( 1,1,\frac{3}{2},\frac{1}{2} \Big)+{}_{2}{{F}_{1}}\Big( 1,1,\frac{3}{2},\frac{1}{4} \Big) \right).
\end{align}
$P_{H_{2}}$ can be resolved by using
\begin{align}
	P_{H_{2}} =
	& A_2\int_{0}^{\infty}P'_{s}(e|\gamma) 
	\Gamma\left(\frac{1}{2}+k_1,B_{\gamma_u} \gamma \right)   d\gamma,
	\nonumber\\=&
	A_2\int_{0}^{\infty} \biggl(
	\frac{1}{2}\sqrt{\frac{\alpha_{h} }{2\pi }}\big( {{H}_{b}}-H_a \big){{\gamma }^{-\frac{1}{2}}}{{e}^{-\frac{\alpha_{h} \gamma }{2}}}-\frac{{{H}_{b}}}{3}\sqrt{\frac{\alpha_{h} }{3\pi }}{{\gamma }^{-\frac{1}{2}}}{{e}^{-\frac{\alpha_{h} \gamma }{3}}}+\frac{{{H}_{b}}}{2}\sqrt{\frac{\alpha_{h} }{6\pi }}{{\gamma }^{-\frac{1}{2}}}{{e}^{-\frac{\alpha_{h} \gamma }{6}}} 
	\biggr)
	\nonumber\\ &\times
	\Gamma\left(\frac{1}{2}+k_1,B_{\gamma_u} \gamma \right)d\gamma
	+A_2\int_{0}^{\infty} \biggl( \frac{2{{H}_{b}}\alpha_{h}}{9\pi }
	{{e}^{-\frac{2\alpha_{h} \gamma }{3}}}{}_{1}{{F}_{1}}\Big( 1,\frac{3}{2},\frac{\alpha_{h}}{3}\gamma \Big) -\frac{{{H}_{b}}\alpha_{h} }{2\sqrt{3}\pi }
	\nonumber\\ &\times
	{{e}^{-\frac{2\alpha_{h} \gamma }{3}}}\left\{{}_{1}{{F}_{1}}\Big( 1,\frac{3}{2},\frac{\alpha_{h}}{2}\gamma \Big)+{}_{1}{{F}_{1}}\Big( 1,\frac{3}{2},\frac{\alpha_{h} }{6}\gamma  \Big) \right\} \biggr)
	\Gamma\left(\frac{1}{2}+k_1,B_{\gamma_u} \gamma \right)d\gamma.
\end{align} 

The above expression can be resolved by using the identity  \cite[eq. 6.455.1, eq.9.14.1]{gradshteyn2014table},  and  given as:
\begin{align}\label{Ph2}
	P_{H_2}&= A_2 \bigg( \alpha^\nu  \bigg(  {\Gamma(\mu+\nu)\over \mu}\bigg( \frac{1}{2}\sqrt{\frac{\alpha_{h} }{2\pi }}\big( {{H}_{b}}-H_a \big) \Bbb{F}_2(\mu,\alpha,\beta_{1})  - \frac{{{H}_{b}}}{3}\sqrt{\frac{\alpha_{h} }{3\pi }} 
	\Bbb{F}_2(\mu,\alpha,\beta_{2})
	\nonumber\\&+
	\frac{{{H}_{b}}}{2}\sqrt{\frac{\alpha_{h} }{6\pi }} \Bbb{F}_2(\mu,\alpha,\beta_{3}) \bigg) + \sum\limits_{n} 
	\frac{ \Gamma(\mu_2+\nu)}{\mu_2 } \Bbb{F}_2(\mu_2,\alpha,\beta_{4})\bigg( \frac{2{{H}_{b}}\alpha_{h}}{9\pi } {\left(\frac{\alpha_{h}}{3}\right)}^{n} -  \frac{{{H}_{b}}\alpha_{h} }{2\sqrt{3}\pi } 
	\nonumber\\&\times
	\left\{ { \left(\frac{\alpha_{h}}{2} \right)}^{n} + { \left(\frac{\alpha_{h}}{6} \right)}^{n}  \right\} \bigg) \bigg)  \bigg).  
\end{align}
On substituting \eqref{Ph1} and \eqref{Ph2} in \eqref{eq:HQ}, to get the closed-form expression as given in \eqref{HGE}.
\end{proof}
\section{Proof of Lemma  \ref{l4}}
\begin{proof}\label{R}
ASER expression for the RQAM scheme is also derived by using the CDF approach. The first order derivative of conditional SEP of RQAM scheme is  	 derived by following the similar approach as in Appendix (\ref{H}) as
\begin{align}\label{Rder}
	\mathrm{P}_{RQAM}^{'}(e|\gamma )=&\frac{a_r{R_1}( {R_2}-1 )}{\sqrt{2\pi \gamma }}{{e}^{-A_r\gamma }}+\frac{b_r {R_2}( {R_1}-1)}{\sqrt{2\pi \gamma }}{{e}^{-B_r\gamma }}-\frac{a_r b_r {R_1}{R_2}}{\pi}{{e}^{-\frac{( {{a_r }^{2}}+{{b_r }^{2}} )\gamma }{2}}}\nonumber\\&\times\bigg\{ {}_{1}{{F}_{1}}\Big( 1,\frac{3}{2},B_r\gamma  \Big)+{}_{1}{{F}_{1}}\Big( 1,\frac{3}{2},A_r \gamma  \Big) \bigg\}.
\end{align}
Further, substituting $P_s^{'}(e|\gamma)$ and $F_{\gamma_{e2e}}(\gamma)$ from (\ref{Rder}) and (\ref{CDF12}), respectively into (\ref{SEP}), we get
\begin{align}
	{{P}_{s}}^{RQAM}=&-\int_{0}^{\infty}P'_{RQAM}(e|\gamma)F_{\gamma_{e2e}}(\gamma) d\gamma,  
	%
	%
	\nonumber\\=&
	-\int_{0}^{\infty}P'_{RQAM}(e|\gamma)d\gamma	
	+
	A_2\int_{0}^{\infty}P'_{RQAM}(e|\gamma) 
	\Gamma\left(\frac{1}{2}+k_1,B_{\gamma_u} \gamma \right)   d\gamma, \nonumber\\
	=& -P_{R_{1}} + P_{R_{2}}, \label{eq:RQ}
\end{align}
$P_{R_1}$ can be resolved by using the identities \cite[eq. 3.351.3,  eq. 7.522.9]{gradshteyn2014table} as
\begin{align}\label{R1}
	P_{R_1}&=a_r{R_1}( {R_2}-1 ) + b_r {R_2}( {R_1}-1)  -\frac{a_r b_r {R_1}{R_2}}{\pi}  \beta_3^{-1}  
	\bigg\{ {}_{1}{{F}_{1}}\left( 1,\frac{3}{2},{B_r \over   \beta_3  }   \right )
	+{}_{1}{{F}_{1}}\left(1, 1,\frac{3}{2},{A_r \over  \beta_3  }  \right ) \bigg\}.
\end{align}	
Similarly, $P_{R_2}$ is resolved by using the identities \cite[eq. 9.14.1, eq. 6.455.1]{gradshteyn2014table} as
\begin{align}\label{R2}
	P_{R_2}&= A_2 \bigg( \alpha^\nu  \bigg(  {\Gamma(\mu+\nu)\over \mu}\bigg(  \frac{a_r{R_1}( {R_2}-1 )}{\sqrt{2\pi  }}\Bbb{F}_2(\mu,\alpha,\beta_{1}) + \frac{b_r {R_2}( {R_1}-1)}{\sqrt{2\pi}}
	\Bbb{F}_2(\mu,\alpha,\beta_{2})\bigg)
	\nonumber\\&
	-
	\sum\limits_{n}  \frac{a_r b_r {R_1}{R_2}}{\pi}  	\frac{ \Gamma(\mu_2+\nu)}{\mu_2}
	\Bbb{F}_2(\mu_2,\alpha,\beta_{3})  \left(  B_r^{n} + A_r^{n}  \right)  \bigg)  \bigg). 
\end{align}
On substituting \eqref{R1} and \eqref{R2} in \eqref{eq:RQ} to get the  generalized ASER expression for RQAM as given in \eqref{RGE}.
\end{proof}
\section{Proof of Lemma  \ref{l5}}
\begin{proof}\label{X}
First order derivative of conditional SEP expression (\ref{Xder}) for XQAM is derived by following the similar approach as in Appendix (\ref{H}) and using the identity  \cite[(9.14.1)]{gradshteyn2014table} and is given as
\begin{align}\label{Xder}
	\mathrm{P}_{X}^{'}\big( e|\gamma  \big)&=\bigg(\frac{\mathbb{A}_X}{2\sqrt{\pi \alpha_{x}}}  {{e}^{-\frac{\gamma}{\alpha_{x}} }} 
	-\frac{A_{x_2}}{\sqrt{\pi\alpha_{x}}} e^{-\frac{ A_{x_1}^2 \gamma}{\alpha_{x}}} \bigg)\gamma^{\frac{-1}{2}}
	+\sum\limits_{n} \gamma^{n}
	\bigg\{ \frac{-16}{M_{x}N_{x}} \sum\limits_{l}  \bigg\{\frac{l}{\pi \alpha_{x}}e^{-\gamma A_{x_3}} \bigg( { \left( \frac{1}{\alpha_{x}} \right) }^{n}  
	\nonumber\\&
	+ {\left( \frac{4l^2}{\alpha_{x}}  \right)}^{n}  \bigg\}
	-\frac{2A_{x_2}}{\pi\alpha_{x}}
	{{e}^{-\gamma\frac{(1+A_{x_1}^2)}{\alpha_{x}} }} 
	\left( { \left( \frac{1}{\alpha_{x}} \right) }^{n}  + {\left( \frac{A_{x_1}^2}{\alpha_{x}}  \right)}^{n}  \right)
	-\frac{k_x}{\pi\alpha_{x}}e^{\frac{-2\gamma}{\alpha_{x}}} { \left( \frac{1}{\alpha_{x}} \right) }^{n} \bigg\}.
\end{align} 
where $\mathbb{A}_X=\left(  -A_n  +k_x+ \frac{4}{M_{x}N_{x}}\left(2\sum_{l=1}^{\frac{ A_{x_1}}{2}-1}  +1 \right)  \right)$.	
The generalized closed-form ASER expression of XQAM can be obtained by substituting $ \mathrm{P}_{s}^{'}\big( e|\gamma  \big)$ and $F_{\gamma_{e2e}}(\gamma)$ from (\ref{Xder}) and (\ref{CDF12}), respectively in (\ref{SEP}) and is given as
\begin{align}
	{{P}_{s}}^{X}=&-\int_{0}^{\infty}P'_{X}(e|\gamma)F_{\gamma_{e2e}}(\gamma) d\gamma,  
	%
	%
	\nonumber\\=&
	-\int_{0}^{\infty}P'_{X}(e|\gamma)d\gamma	
	+
	A_2\int_{0}^{\infty}P'_{X}(e|\gamma) 
	\Gamma\left(\frac{1}{2}+k_1,B_{\gamma_u} \gamma \right)   d\gamma
	= -P_{X_{1}} + P_{X_{2}}, \label{eq:X}
\end{align}
$P_{X_{1}} = \int_{0}^{\infty}P'_{X}(e|\gamma)d\gamma$ can be resolved by using the identity \cite[eq. 3.351.3]{gradshteyn2014table} as
\begin{align}\label{Px1}
	P_{X_{1}} &=  \frac{-1}{2} \mathbb{A}_X +\frac{2}{{M_{x}N_{x}}} 
	+ \frac{16}{M_{x}N_{x}} \sum\limits_{l} \frac{l}{\pi \alpha_{x} } A_{x_3}^{-1} 
	\left( {}_{1}{{F}_{1}}\Big( 1,\frac{3}{2},\frac{1}{A_{x_3} \alpha_{x}} \Big) + {}_{1}{{F}_{1}}\Big( 1,\frac{3}{2},\frac{4 l^2}{A_{x_3} \alpha_{x}} \Big) \right)
	\nonumber\\&
	-2\frac{A_{x_2}}{\pi\alpha_{x}} \beta_4^{-1} \bigg( {}_{1}{{F}_{1}}\Big( 1,\frac{3}{2},\frac{A_{x_1}^2}{\beta_4 \alpha_{x}} \Big) +{}_{1}{{F}_{1}}\Big( 1,\frac{3}{2},\frac{1}{\beta_4 \alpha_{x}} \Big) \bigg)
	+\frac{k_x}{\pi\alpha_{x}} {\beta_5}^{-1}{}_{1}{{F}_{1}}\Big( 1,\frac{3}{2},\frac{1}{\beta_5 \alpha_{x}} \Big)   \bigg).
\end{align}
$	P_{X_{2}} $ can be obtained by using the identity  \cite[eq. 6.455.1]{gradshteyn2014table}  as
\begin{align} \label{Px2}
	P_{X_2}&= A_2 \bigg( \alpha^\nu  \bigg(  {\Gamma(\mu+\nu)\over \mu}\bigg( \frac{1}{2\sqrt{\pi \alpha_{x}}}\mathbb{A}_X \Bbb{F}_2(\mu,\alpha,\beta_{1}) 
	-\frac{A_{x_2}}{\sqrt{\pi\alpha_{x}}} \Bbb{F}_2(\mu,\alpha,\beta_{2})  \bigg)  
	+ \sum\limits_{n}  \frac{ \Gamma(\mu_2+\nu)}{\mu_2} 
	\nonumber\\&\times
	\bigg(   \frac{-16}{M_{x}N_{x}} \sum\limits_{l} \frac{l}{\pi \alpha_{x}}
	\Bbb{F}_2(\mu_2,\alpha,\beta_{3})  \left( { \left( \frac{1}{\alpha_{x}} \right) }^{n}  + {\left( \frac{4l^2}{\alpha_{x}}  \right)}^{n}  \right) - 2\frac{A_{x_2}}{\pi\alpha_{x}} \left( { \left( \frac{1}{\alpha_{x}} \right) }^{n}  + {\left( \frac{A_{x_1}^2}{\alpha_{x}}  \right)}^{n}  \right) 
	\nonumber\\&\times
	\Bbb{F}_2(\mu_2,\alpha,\beta_{4})  	
	-\frac{k_x}{\pi\alpha_{x}} \Bbb{F}_2(\mu_2,\alpha,\beta_{5}) { \left( \frac{1}{\alpha_{x}} \right) }^{n}  \bigg) \bigg)   \bigg). 
\end{align}

On substituting \eqref{Px1} and \eqref{Px2} in \eqref{eq:X}, the generalized ASER expression for XQAM is obtained as in \eqref{XGE}.
\end{proof}
\section{Proof of Lemma  \ref{l6}}
\begin{proof}\label{ER1}
On substituting \eqref{CDF12} in \eqref{Er}, the integral is given as  
\begin{align}
	C_R &=  {1\over{2\ln2}} \int_{0}^{\infty} {  1- \left( 1- A_2   \Gamma\left(\frac{1}{2}+k_1,B_{\gamma_u} \gamma \right)   \right) \over  1+\gamma }  d\gamma 
	= {A_2\over{2\ln2}} \int_{0}^{\infty} {   \Gamma\left(\frac{1}{2}+k_1,B_{\gamma_u} \gamma \right)  \over  1+\gamma }  d\gamma. \label{eq:2}
\end{align}
The exact analysis of the above integral is not possible and hence the $\Gamma(a,b)$ can be represented in terms of Meijer-G function by using the identity given in \cite[eq. 8.4.16.2]{prudnikov1989integrals}. \eqref{eq:2} can be re-written as 
\begin{align}
	C_{R_ 2} &= {A_2\over{2\ln2}} \int_{0}^{\infty} { G_{1,2}^{2,0} 1 \over 1+\gamma}\left( \begin{matrix}
		\beta \gamma
	\end{matrix} \bigg| \begin{matrix}
		1 \\ 0,\alpha
	\end{matrix} \right) d\gamma, \label{er1}
\end{align}
where $\beta= \frac { 1} {2{\bar {\gamma }}\sigma_{\gamma}^2 } $ and $\alpha=\frac{1}{2}+k_1$. The above integral can be resolved by using the identity \cite[eq. 7.8.11.5]{gradshteyn2014table} to get the closed-form ergodic rate expression for $S \rightarrow H_I \rightarrow U $ link as in \eqref{Er}
\end{proof}
\section{Proof of Lemma  \ref{l7}}
\begin{proof} \label{OP2}
The closed-form outage probability expression can be obtained by derived the CDFs corresponding to $\gamma_s$ and $\gamma_u$. $\gamma_s$, is the instantaneous SNR corresponds to the $S \rightarrow R$ link which is Rician fading channel. The CDF of $\gamma_s$ is given as \cite[eq. 2.3-57]{proakis1995digital}
\begin{align}\label{F1}
	F_{\gamma_s}(\gamma_{th})= 1-Q_{1}\left ({ \frac { \mu_{\gamma_s} } {\sigma_{\gamma_s} }, \frac {\gamma_{th}}{\sigma_{\gamma_s} } }\right), \quad  {\gamma_{th}} > 0		
\end{align}
$\sigma_{\gamma_s}=\sqrt{\frac{\mu_{\gamma_s}^2}{2 K}}$ \cite[eq. 2.3-60]{proakis1995digital}.  In \eqref{a2}, $\gamma_u$ is the sum of products of $\lambda_i$ and  $\kappa_i$ which are independent shadowed Rician  and Nakagami-$m$ fading channels, respectively. A similar approach is followed to characterize the PDF and CDF of $\gamma_u $ as in Appendix A. Lets consider $\gamma_u=\bar{\gamma_u }A^2$, where in  $A=\sum_{i=0}^{N}\tilde{A}$
Let $\tilde{A}=BC$ where $B$ and $C$ are independent random variables.
The mean and variances of $\tilde{A}$  are given as $\mu_A=\mathrm{E}(A)=\sum_{i=1}^{N}\mathrm{E}(\tilde{A}_i)=N\mathrm{E}(\tilde{A})$ and $\sigma_A^2=\mathrm{Var}({A})=\sum_{i=1}^{N}\mathrm{Var}(\tilde{A}_i)=N\mathrm{Var}(\tilde{A})$, respectively. Wherein
\begin{align}
	\mathrm{E}(\tilde{A})&=\frac{\Gamma{(m_g+\frac{1}{2})}}{\Gamma{(m_g)}} \left(\frac{{\sigma_g^2}}{m_g} \right)^\frac{1}{2} B_0^{m_h} (2b_0)^{1/2} \Gamma \left( {{3}\over {2}}  \right) {}_2F_1 \left( {{3}\over {2}},\, m_h,\, 1,\, B_1 \right),
\end{align}
\begin{align}
	\mathrm{Var}(\tilde{A})&= \frac{\Gamma{(m_g+1)}}{\Gamma{(m_g)}} \left(\frac{{\sigma_g^2}}{m_g} \right)  B_0^{m_h} (2b_0)  {}_2F_1 \left( 2,\, m_h,\, 1,\, B_1 \right) - \bigg( \frac{\Gamma{(m_g+\frac{1}{2})}}{\Gamma{(m_g)}}	\left(\frac{{\sigma_g^2}}{m_g} \right)^\frac{1}{2} B_0^{m_h}
	\nonumber\\& \times
	(2b_0)^{1/2} \Gamma \left( {{3}\over {2}}  \right) {}_2F_1 \left( {{3}\over {2}},\, m_h,\, 1,\, B_1 \right) \bigg)^2.
\end{align}
Thus, by invoking CLT for sufficiently large number of reflecting meta-surfaces $A^2$ follows a non-central chi-square random variable with one degree of freedom with mean $\mu_A=N\mu_{\tilde{A}}$ and variance $\sigma^2_A=N\sigma^2_{\tilde{A}}$. Hence, the  CDF of  $\gamma_{su}$ is given by \cite[eq. (2.3-35)]{proakis1995digital}
\begin{align} \label{F2}
	F_{\gamma_u}(\gamma_u)=1-Q_{\frac {1}{2}}\left ({\frac { {\mu_{\gamma_u} }}{\sigma_{\gamma_u} },\frac {\sqrt {\gamma_u}}{\sqrt {\bar {\gamma_u }}\sigma_{\gamma_u} }}\right), \quad  {\gamma_u} > 0
\end{align}
where $\mu_{\gamma_u}= \mu_A$ and $\sigma_{\gamma_u}=\sigma_A$. On substituting \eqref{F1} and \eqref{F2} in \eqref{Pout2}, to get the closed-form expression as given in \eqref{CDF2}. 
\end{proof}
\section{Proof of Lemma  \ref{l8}}
\begin{proof} \label{ER2}
On substituting \eqref{CDF21} in \eqref{Er}, the integral is given as  
	\begin{align}
		C_R &=  {1\over{2\ln2}} \int_{0}^{\infty} {  1-\left(1- A_3 \exp[-\Omega  \gamma ] \Gamma({1\over 2}+k_2,\beta_{2}^{2}/2) \gamma^{n} \right) \over  1+\gamma }  d\gamma, 
		\nonumber\\&
		= {A_3\over{2\ln2}} \int_{0}^{\infty} \gamma^{n} \exp[-\Omega  \gamma ] {\Gamma(\frac{1}{2}+k_1,B_{\gamma_u} \gamma ) \over 1+\gamma} d\gamma, \label{eq: er2}
	\end{align}
		%
	%
	By using the identities \cite[eq. 8.4.16.2]{prudnikov1989integrals} and  \cite[eq. 1.211.1]{gradshteyn2014table},  \eqref{eq: er2}, can be re-written  as
	\begin{align}
		C_{R_ 2} &= {A_3\over{2\ln2} } \sum_{j=0}^{\infty} {(-\Omega)^j \over j!} \int_{0}^{\infty}  \gamma^{n+j} 
		(1+\gamma)^{-1} G_{1,2}^{2,0} \left( \begin{matrix}
			\beta \gamma
		\end{matrix} \bigg| \begin{matrix}
			1 \\ 0,\alpha
		\end{matrix} \right)  d\gamma, \label{er3}
	\end{align}
	The above integral is resolved by using the identity \cite[eq. 7.811.5]{gradshteyn2014table} to get the closed expression for ergodic rate of the system model \figurename{ \ref{SY2}} as given in \eqref{ERG2}.
	
	
\end{proof}
\bibliographystyle{IEEEtran}
\bibliography{irs_bibfile}

\begin{thebibliography}{10}
\providecommand{\url}[1]{#1}
\csname url@samestyle\endcsname
\providecommand{\newblock}{\relax}
\providecommand{\bibinfo}[2]{#2}
\providecommand{\BIBentrySTDinterwordspacing}{\spaceskip=0pt\relax}
\providecommand{\BIBentryALTinterwordstretchfactor}{4}
\providecommand{\BIBentryALTinterwordspacing}{\spaceskip=\fontdimen2\font plus
\BIBentryALTinterwordstretchfactor\fontdimen3\font minus
  \fontdimen4\font\relax}
\providecommand{\BIBforeignlanguage}[2]{{%
\expandafter\ifx\csname l@#1\endcsname\relax
\typeout{** WARNING: IEEEtran.bst: No hyphenation pattern has been}%
\typeout{** loaded for the language `#1'. Using the pattern for}%
\typeout{** the default language instead.}%
\else
\language=\csname l@#1\endcsname
\fi
#2}}
\providecommand{\BIBdecl}{\relax}
\BIBdecl

\bibitem{chaoub20216g}
A.~Chaoub, M.~Giordani, B.~Lall, V.~Bhatia, A.~Kliks, L.~Mendes, K.~Rabie,
  H.~Saarnisaari, A.~Singhal, N.~Zhang \emph{et~al.}, ``{6G for bridging the
  digital divide: Wireless connectivity to remote areas},'' \emph{IEEE Wireless
  Commun.}, vol.~29, no.~1, pp. 160--168, Jul. 2021.

\bibitem{giordani2020non}
M.~Giordani and M.~Zorzi, ``{Non-terrestrial networks in the 6G era: Challenges
  and opportunities},'' \emph{IEEE Netw.}, vol.~35, no.~2, pp. 244--251, Dec.
  2020.

\bibitem{huang2019performance}
Q.~Huang, M.~Lin, W.-P. Zhu, S.~Chatzinotas, and M.-S. Alouini, ``{Performance
  analysis of integrated satellite-terrestrial multiantenna relay networks with
  multiuser scheduling},'' \emph{IEEE Trans. Aerospace Electron. Sys.},
  vol.~56, no.~4, pp. 2718--2731, Nov. 2019.

\bibitem{3gpp2019study}
3GPP, ``{Study on New Radio (NR) to support non-terrestrial networks},'' 2020.

\bibitem{ye2022non}
J.~Ye, J.~Qiao, A.~Kammoun, and M.-S. Alouini, ``{Non-terrestrial
  communications assisted by reconfigurable intelligent surfaces},''
  \emph{Proc. IEEE}, May 2022.

\bibitem{liu2018space}
J.~Liu, Y.~Shi, Z.~M. Fadlullah, and N.~Kato, ``{Space-air-ground integrated
  network: A survey},'' \emph{IEEE Commun. Surv. Tuts.}, vol.~20, no.~4, pp.
  2714--2741, May. 2018.

\bibitem{holloway2012overview}
C.~L. Holloway, E.~F. Kuester, J.~A. Gordon, J.~O'Hara, J.~Booth, and D.~R.
  Smith, ``{An overview of the theory and applications of metasurfaces: The
  two-dimensional equivalents of metamaterials},'' \emph{IEEE Antennas Propag.
  Mag.}, vol.~54, no.~2, pp. 10--35, Apr. 2012.

\bibitem{alghamdi2020intelligent}
R.~Alghamdi, R.~Alhadrami, D.~Alhothali, H.~Almorad, A.~Faisal, S.~Helal,
  R.~Shalabi, R.~Asfour, N.~Hammad, A.~Shams \emph{et~al.}, ``{Intelligent
  surfaces for 6G wireless networks: A survey of optimization and performance
  analysis techniques},'' \emph{IEEE Access}, Oct. 2020.

\bibitem{tekbiyik2020reconfigurable}
K.~Tekb{\i}y{\i}k, G.~K. Kurt, A.~R. Ekti, A.~G{\"o}r{\c{c}}in, and
  H.~Yanikomeroglu, ``{Reconfigurable intelligent surfaces empowered THz
  communication in LEO satellite networks},'' \emph{arXiv preprint
  arXiv:2007.04281}, Jul. 2020.

\bibitem{xu2021intelligent}
S.~Xu, J.~Liu, Y.~Cao, J.~Li, and Y.~Zhang, ``{Intelligent reflecting surface
  enabled secure cooperative transmission for satellite-terrestrial integrated
  networks},'' \emph{IEEE Trans. Veh. Technol.}, vol.~70, no.~2, pp.
  2007--2011, Feb. 2021.

\bibitem{li2021reconfigurable}
J.~Li, S.~Xu, J.~Liu, Y.~Cao, and W.~Gao, ``{Reconfigurable intelligent surface
  enhanced secure aerial-ground communication},'' \emph{IEEE Trans. Commun.},
  vol.~69, no.~9, pp. 6185--6197, Jun. 2021.

\bibitem{ibrahim2021exact}
H.~Ibrahim, H.~Tabassum, and U.~T. Nguyen, ``{Exact coverage analysis of
  intelligent reflecting surfaces with Nakagami-m channels},'' \emph{IEEE
  Trans. Veh. Technol.}, vol.~70, no.~1, pp. 1072--1076, Jan. 2021.

\bibitem{tian2022enabling}
X.~Tian, N.~Gonzalez-Prelcic, and T.~Shimizu, ``{Enabling NLoS LEO Satellite
  Communications with Reconfigurable Intelligent Surfaces},'' \emph{arXiv
  preprint arXiv:2205.15528}, 2022.

\bibitem{alfattani2021aerial}
S.~Alfattani, W.~Jaafar, Y.~Hmamouche, H.~Yanikomeroglu, A.~Yonga{\c{c}}oglu,
  N.~D. {\DJ}{\`a}o, and P.~Zhu, ``{Aerial platforms with reconfigurable smart
  surfaces for 5G and beyond},'' \emph{IEEE Commun. Mag.}, vol.~59, no.~1, pp.
  96--102, Feb. 2021.

\bibitem{bariah2021ris}
L.~Bariah, L.~Mohjazi, H.~Abumarshoud, B.~Selim, M.~Tatipamula, M.~A. Imran,
  H.~Haas \emph{et~al.}, ``{RIS-Assisted Space-Air-Ground Integrated Networks:
  New Horizons for Flexible Access and Connectivity},'' 2021.

\bibitem{ramezani2022toward}
P.~Ramezani, B.~Lyu, and A.~Jamalipour, ``{Toward RIS-Enhanced Integrated
  Terrestrial/Non-Terrestrial Connectivity in 6G-Enabled IoE Era},''
  \emph{arXiv preprint arXiv:2203.11312}, 2022.

\bibitem{tekbiyik2022reconfigurable}
K.~Tekbiyik, G.~Kurt, A.~Ekti, and H.~Yanikomeroglu, ``{Reconfigurable
  Intelligent Surfaces in Action: For Nonterrestrial Networks: Employing
  Reconfigurable Intelligent Surfaces},'' \emph{IEEE Vehicular Technol. Mag.},
  2022.

\bibitem{bjornson2019intelligent}
E.~Bj{\"o}rnson, {\"O}.~{\"O}zdogan, and E.~G. Larsson, ``{Intelligent
  reflecting surface versus decode-and-forward: How large surfaces are needed
  to beat relaying?}'' \emph{IEEE Wireless Commun. Lett.}, vol.~9, no.~2, pp.
  244--248, Oct. 2019.

\bibitem{boulogeorgos2020performance}
A.-A.~A. Boulogeorgos and A.~Alexiou, ``{Performance analysis of reconfigurable
  intelligent surface-assisted wireless systems and comparison with
  relaying},'' \emph{IEEE Access}, vol.~8, pp. 94\,463--94\,483, May 2020.

\bibitem{di2020reconfigurable}
M.~Di~Renzo, K.~Ntontin, J.~Song, F.~H. Danufane, X.~Qian, F.~Lazarakis,
  J.~De~Rosny, D.-T. Phan-Huy, O.~Simeone, R.~Zhang \emph{et~al.},
  ``{Reconfigurable intelligent surfaces vs. relaying: Differences,
  similarities, and performance comparison},'' \emph{IEEE Open J. Commun.
  Soc.}, vol.~1, pp. 798--807, Jun. 2020.

\bibitem{sikri2021reconfigurable}
A.~Sikri, A.~Mathur, P.~Saxena, M.~R. Bhatnagar, and G.~Kaddoum,
  ``{Reconfigurable intelligent surface for mixed FSO-RF systems with
  co-channel interference},'' \emph{IEEE Commun. Lett.}, vol.~25, no.~5, pp.
  1605--1609, Feb. 2021.

\bibitem{dolas2022performance}
K.~Dolas and M.~R. Bhatnagar, ``{On Performance of IRS-Assisted Hybrid
  Satellite-Terrestrial Cooperative Communication},'' \emph{IEEE Trans.
  Aerospace Electron. Sys.}, Aug. 2022.

\bibitem{dong2022intelligent}
H.~Dong, C.~Hua, L.~Liu, W.~Xu, and R.~Tafazolli, ``{Intelligent Reflecting
  Surface-Aided Integrated Terrestrial-Satellite Networks},'' \emph{IEEE Trans.
  Wireless Commun.}, Oct. 2022.

\bibitem{singya2021survey}
P.~K. Singya, P.~Shaik, N.~Kumar, V.~Bhatia, and M.-S. Alouini, ``{A Survey on
  Higher-Order QAM Constellations: Technical Challenges, Recent Advances, and
  Future Trends},'' \emph{IEEE Open J. Commun. Soc.}, vol.~2, pp. 617--655,
  Mar. 2021.

\bibitem{Parvez_D2D_Access_2019}
S.~{Parvez}, P.~K. {Singya}, and V.~{Bhatia}, ``On {ASER} analysis of energy
  efficient modulation schemes for a device-to-device {MIMO} relay network,''
  \emph{IEEE Access}, vol.~8, pp. 2499--2512, Dec. 2019.

\bibitem{parvez2019impact}
S.~Parvez, P.~K. Singya, and V.~Bhatia, ``On impact of imperfect {CSI} over
  hexagonal {QAM} for {TAS/MRC-MIMO} cooperative relay network,'' \emph{IEEE
  Commun. Lett.}, vol.~23, no.~10, pp. 1721--1724, Jul. 2019.

\bibitem{shaik2019performance}
P.~Shaik, P.~K. Singya, and V.~Bhatia, ``{Performance analysis of QAM schemes
  for non-regenerative cooperative MIMO network with transmit antenna
  selection},'' \emph{AEU-Int. J. Electron. Commun.}, vol. 107, pp. 298--306,
  Jul. 2019.

\bibitem{garg2019performance}
K.~K. Garg, P.~Singya, and V.~Bhatia, ``{Performance analysis of NLOS
  ultraviolet communications with correlated branches over turbulent
  channels},'' \emph{J. Opt. Commun. Netw.}, vol.~11, no.~11, pp. 525--535,
  Nov. 2019.

\bibitem{10.1117/1.OE.59.1.016106}
Shaik, K.~K. Garg, and V.~Bhatia, ``{On impact of imperfect channel state
  information on dual-hop nonline-of-sight ultraviolet communication over
  turbulent channel},'' \emph{Opt. Eng.}, vol.~59, no.~1, pp. 1 -- 14, Jan.
  2020.

\bibitem{singya2021performance}
P.~K. Singya and M.-S. Alouini, ``Performance of uav assisted multiuser
  terrestrial-satellite communication system over mixed fso/rf channels,''
  \emph{IEEE Trans. Aeros. Electron. Syst.}, Sep.

\bibitem{garg2022performance}
K.~K. Garg, P.~Shaik, V.~Bhatia, and O.~Krejcar, ``{On the performance of a
  relay assisted hybrid RF-NLOS UVC system with imperfect channel
  estimation},'' \emph{J. Optical Commun. Netw.}, vol.~14, no.~4, pp. 177--189,
  2022.

\bibitem{abdi2003new}
A.~Abdi, W.~C. Lau, M.-S. Alouini, and M.~Kaveh, ``{A new simple model for land
  mobile satellite channels: First-and second-order statistics},'' \emph{IEEE
  Trans. Wireless Commun.}, vol.~2, no.~3, pp. 519--528, May 2003.

\bibitem{suraweera2009performance}
H.~A. Suraweera, G.~K. Karagiannidis, and P.~J. Smith, ``{Performance analysis
  of the dual-hop asymmetric fading channel},'' \emph{IEEE Transactions on
  Wireless Communications}, vol.~8, no.~6, pp. 2783--2788, Jun.

\bibitem{proakis1995digital}
J.~G. Proakis and M.~Salehi, ``{Digital Communications, McGraw-Hill},''
  \emph{Inc., New York,}, 1995.

\bibitem{shuai2022transmit}
H.~Shuai, K.~Guo, K.~An, Y.~Huang, and S.~Zhu, ``{Transmit Antenna Selection in
  NOMA-based Integrated Satellite-HAP-Terrestrial Networks with Imperfect CSI
  and SIC},'' \emph{IEEE Wireless Commun. Lett.}, Apr. 2022.

\bibitem{basar2019wireless}
E.~Basar, M.~Di~Renzo, J.~De~Rosny, M.~Debbah, M.-S. Alouini, and R.~Zhang,
  ``Wireless communications through reconfigurable intelligent surfaces,''
  \emph{IEEE Access}, vol.~7, pp. 116\,753--116\,773, Aug. 2019.

\bibitem{parvez2019aser}
S.~Parvez, P.~K. Singya, V.~Bhatia, and N.~Kumar, ``Aser analysis of cross qam
  for tas/mrc-mimo cooperative relay system with imperfect csi,'' in \emph{IEEE
  Adv. Net. Telecommun. Syst. (ANTS)}.\hskip 1em plus 0.5em minus 0.4em\relax
  IEEE, 2019, pp. 1--7.

\bibitem{Garg_2021}
K.~Garg, P.~Shaik, V.~Bhatia, and O.~Krejcar, ``On the performance of relay
  assisted hybrid {RF}-{NLOS} {UVC} sy.stem with imperfect channel
  estimation,'' \emph{J. Optical Commun. Netw.}, vol.~14, no.~3, Dec. 2021.

\bibitem{SADHWANI201763}
D.~Sadhwani and R.~N. Yadav, ``{A simplified exact expression of SEP for cross
  QAM in AWGN channel from M$\times$N rectangular QAM and its usefulness in
  Nakagami-m fading channel},'' \emph{AEU - International Journal of
  Electronics and Communications}, vol.~74, pp. 63 -- 74, 2017.

\bibitem{annamalai2009new}
A.~Annamalai, C.~Tellambura, and J.~Matyjas, ``{A new twist on the generalized
  Marcum Q-function QM (a, b) with fractional-order M and its applications},''
  in \emph{2009 6th IEEE Consumer Communications and Networking
  Conference}.\hskip 1em plus 0.5em minus 0.4em\relax IEEE, Jan. 2009, pp.
  1--5.

\bibitem{gradshteyn2014table}
I.~S. Gradshteyn and I.~M. Ryzhik, \emph{{Table of integrals, series, and
  products}}.\hskip 1em plus 0.5em minus 0.4em\relax Academic Press, 2014.

\bibitem{3gpp}
{3GPP}, ``{Universal Terrestrial Radio Access (UTRA): repeater planning
  guidelines and system analysis},'' \emph{3rd Generation Partnership Project
  (3GPP), TR 25.956 V16.0.0}, Jun. 2020.

\bibitem{xing2021high}
Y.~Xing, F.~Hsieh, A.~Ghosh, and T.~S. Rappaport, ``{High altitude platform
  stations (HAPS): Architecture and system performance},'' in \emph{2021 IEEE
  93rd Vehicular Technology Conference (VTC2021-Spring)}.\hskip 1em plus 0.5em
  minus 0.4em\relax IEEE, Apr. 2021, pp. 1--6.

\bibitem{abramowitz1964handbook}
M.~Abramowitz and I.~A. Stegun, \emph{{Handbook of mathematical functions: with
  formulas, graphs, and mathematical tables}}.\hskip 1em plus 0.5em minus
  0.4em\relax 9th ed. {N}ew {Y}ork, {NY}, {USA}: Dover, 1970.

\bibitem{prudnikov1989integrals}
A.~P. Prudnikov, J.~A. Bry{\v{c}}kov, and O.~I. Mari{\v{c}}ev, \emph{{Integrals
  and series. Vol. 3, More special functions}}.\hskip 1em plus 0.5em minus
  0.4em\relax New York: Gordon and Breach, 1989.

\end{thebibliography}
\end{document}